\documentclass{fundam}%

\usepackage{amsmath}
\usepackage{amsfonts}
\usepackage{amssymb}
\usepackage{graphicx}%

\begin{document}
	
\setcounter{page}{89}
\publyear{22}
\papernumber{2120}
\volume{186}
\issue{1-4}

     \finalVersionForARXIV
    %% \finalVersionForIOS

\title{Order-theoretic Trees: \\
              Monadic Second-order Descriptions and Regularity}

\author{Bruno Courcelle\thanks{Address of correspondence:  LaBRI, Universit\'e de Bordeaux, 351
                            Cours de la Lib\'eration, 33400, Talence, France}
 \\
Bordeaux University, CNRS, LaBRI, France\\
courcell@labri.fr}

\runninghead{B. Courcelle}{Order-theoretic Trees: Monadic Second-order Descriptions and Regularity}

\maketitle

\vspace*{-6mm}
\begin{abstract}
 An \emph{order-theoretic forest} is a countable partial
order such that the set of elements larger than any element is linearly
ordered.\ It is an \emph{order-theoretic tree} if any two elements have an
upper-bound.\ The order type of a \emph{branch} (a maximal linearly ordered
subset) can be any countable linear order.\ Such generalized infinite trees
yield convenient definitions of the rank-width and the modular decomposition
of countable graphs.

 We define an algebra based on only four operations that generate up to
isomorphism and \emph{via} infinite terms these order-theoretic trees and
forests.\ We prove that the associated \emph{regular} objects,\emph{ i.e.,}
those defined by regular terms,\ are exactly the ones that are the unique
models of monadic second-order sentences.\

 We adapt some tools that we have used in a previous article for proving
a similar result for \emph{join-trees}, \emph{i.e.}, for order-theoretic trees
such that any two nodes have a least upper-bound.

\end{abstract}

\section{Introduction}

Countably infinite trees have been studied in Fundamental Computer Science for
many purposes: semantics of programs \cite{Arn,CouHdbk9}, theory of games
\cite{BCJ,Tho90,VW} and distibuted computing \cite{Ang,CouMet}, just to name a
few topics.\ To be usable for obtaining algorithms, these trees must have
finitary descriptions: several notions have been developped for this purpose,
\emph{e.g.}, automata of various types, logical descriptions and equation
systems \cite{Blu,Cou83}.

In the logical setting of the \emph{Theory of Relations} \cite{Fra}, a more
general notion of tree is used : an \emph{order-theoretic forest} (an
\emph{O-forest} in short) is a countable partial order whose elements are
called \emph{nodes} and such that the set of nodes larger than any node is
linearly ordered; it is an \emph{order-theoretic tree} (an \emph{O-tree}) if
any two nodes have an upper-bound and a \emph{join-tree} if any two nodes have
a least upper-bound. The order type of a linearly ordered subset can be any,
possibly dense, linear order.\ Join-trees yield convenient definitions of the
rank-width and the modular decomposition of countable graphs
\cite{Cou14,CouDel}.

\medskip
Finitary descriptions of these objects can be given by logical sentences.\ For
an example, the ordered set of rational numbers is, up to isomorphism, the
unique linearly ordered countable set that is dense without maximal and
minimal elements.\ This characterization is first-order (FO)
expressible.\ First-order logic is actually not powerful enough for
characterizing the structures we are interested in.\ Monadic second-order
(MSO) logic is much more expressive, and furthermore, its decidability on
infinite trees is a fundamental result due to Rabin (see, \emph{e.g.}
\cite{Tho90}).\ This fundamental result is based on an equivalence between
certain finite automata and MSO descriptions.\ Such an equivalence was proved
for finite words by Trakhtenbrot \cite{Tra} and, independently, by others for
finite trees. MSO\ formulas can also express transformations of structures
that are very useful in proofs.

The rooted trees used in semantics are, in many cases, infinite terms (like
are formal power series) built with finite sets of function symbols. We call
them $F$-\emph{terms}, where $F$ is the set of function symbolds: they can be
conveniently handled as labelled rooted trees, and we will discuss them by
using notions concerning trees: nodes, root, ancestor etc.

An $F$-term is \emph{regular} if it has finitely many different subterms,
equivalently if it is a component of the unique solution of a finite equation
system of a certain type. Such systems yield finitary descriptions of regular
$F$-terms. More complex equation systems yield wider notions of $F$-terms
\cite{Blu,Cou83} whose MSO-theory is decidable, so that some of our results
are applicable to them.

The existence of equivalent characterizations of a same class of objects
indicates a certain \emph{robustness} as this class does not depend on a
choice of definitions that may sound arbitrary.\ In particular, an $F$-term is
regular if and only if it is the unique model of an MSO sentence.\ (This
characterization uses a description of $F$-terms by logical structures).

Our objective is to obtain finitary descriptions of certain O-trees and
O-forests, in particular by regular terms. For this purpose, we define an
algebra structure on the class of O-forests that uses only three
operations\footnote{An additional constant denotes the empty O-forest.}%
.\ These operations can generate all O-trees up to isomorphism \emph{via}
infinite terms. The \emph{regular O-trees} are those defined by \emph{regular
}$\emph{F}$\emph{-terms}.\ Our main theorem states that an O-tree is regular
if and only if it is \emph{monadic second-order definable}, \emph{i.e.}, is
the unique model (up to isomorphism) of an MSO sentence. In this way, we
extend the corresponding result obtained for join-trees in \cite{CouLMCS}. The
proof uses the corresponding result for \emph{arrangements} \cite{Tho86}%
.\ Arrangements are labelled linear orders, hence are generalized words whose
ordered sets of positions may be dense.

A main technical tool is the notion of \emph{structuring} of an O-forest : it
is a partition in convex linearly ordered subsets that form a kind of tree.\ A
regular O-forest has a structuring consisting of regular, whence
MSO-definable, arrangements; furthermore, the tree of these arrangements is in
some sense regular. These notions are collected in a finitary\emph{ regular
description scheme}. A regular O-tree is definable, equivalently, by a regular
$F$-term, by a regular description scheme and by an MSO sentence that can use
a finiteness set predicate. (Such a use is necessary).

All constructions are effective (although intractable)\ and so, the
isomorphism of two regular O-trees is decidable.

In Section 2\ we review definitions and results concerning partial orders,
rooted trees, $F$-terms, arrangements and monadic second-order logic.\ In
Section 3, we define O-forests and O-trees, their structurings and
descriptions schemes. In Section 4 we define the algebra of structured
O-forests.\ We get the notion of a regular O-forest and we prove the main theorem.

\section{Definitions and a few lemmas}

In the present article, all ordered sets, trees and logical structures are
\emph{countable}, which means finite or countably infinite. We denote by
$X\uplus Y$ the union of sets $X$ and $Y$ if we want to insist that they are
disjoint.\ Isomorphism of ordered sets, trees and other logical structures is
denoted by $\simeq$. The restriction of a relation $R$ or a function $f$
defined on a set $V$\ to a subset $W$ is denoted by $R\upharpoonright W$ or
$f\upharpoonright W$ respectively.

\subsection{Orders}

\textbf{Notation and definitions 2.1\ }

For partial orders $\leq,\sqsubseteq$, ... we denote respectively by
$<,\sqsubset$, ... the corresponding strict partial orders. In many cases a
partial order $\leq$ can be defined in a short way from the corresponding
strict partial order $<$ so that $x\leq y$ holds if and only if $x=y$ or $x<y$.

\medskip
Let $(V,\leq)$ be a partial order.\ For subsets $X,Y$ of $V$, $X<Y$ means that
$x<y$ for every $x\in X$ and $y\in Y$.\ We write $X<y$ instead of $X<\{y\}$
and similarly for $x<Y$. The least upper bound of $x$ and $y$ is denoted by
$x\sqcup y$ if it exists and is called their \emph{join}.\ The notation $x\bot
y$ means that $x$ and $y$ are incomparable.\ A subset $Y$ of $V$\ is
\emph{convex} if $y\in Y$ whenever $x,z\in Y$\ and $x<y<z$. A
\emph{line\footnote{In \cite{Cou14} we called \emph{line} a linearly ordered
subset, without imposing the convexity property. The present definition is
from \cite{CouLMCS}}\ }is a convex subset of $V$ that is linearly
ordered.\ Particular\ notations for convex sets (that are not necessarly
linearly ordered) are $[x,y]$ denoting $\{z\mid x\leq z\leq y\}$%
,$\ ]-\infty,x[$ denoting $\{y\mid y<x\}$ (even if $V$ is finite),
$[x,+\infty\lbrack$ denoting $\{y\mid x\leq y\}.$ In a linearly ordered set, a
line is called an \emph{interval}.

\medskip
Let $(N,\leq)$\ and $(N^{\prime},\leq^{\prime})$ be partial orders.\ An
\emph{embedding} $j:(N,\leq)\rightarrow(N^{\prime},$ $\leq^{\prime})$\ is an
injective order-preserving map such that $x\leq y$ if and only if
$j(x)\leq^{\prime}j(y)$; in this case,\ $(N,\leq)$\ is isomorphic by $j$ to
$(j(N),\leq^{\prime}\upharpoonright j(N))$, a suborder of $(N^{\prime}%
,\leq^{\prime})$.\ We say that $j$ is a \emph{join-embedding} if, furthermore,
$j(x\sqcup y)=j(x)\sqcup j(y)$ whenever $x\sqcup y$ is defined.

The first infinite ordinal and the linear order$\ (\mathbb{N},\leq)$ are
denoted by $\omega$.

\bigskip
\noindent \textbf{Definitions 2.2 }: \emph{Cuts in linear orders.}\

(a) \ A (Dedekind) \emph{cut }of a linear order $(U,\leq)$ is a pair
$(U_{1},U_{2})$ of non-empty intervals such that $U=U_{1}\uplus U_{2}$\ and
$U_{1}<U_{2}$. For an example, an element $x\in U$ that is not maximal defines
the cut $(]-\infty,x],]x,+\infty\lbrack)$.

(b) If $K$ is a set of cuts, we extend $\leq$\ into a linear order on $U\uplus
K,$ also denoted by $\leq,$ such that:

\begin{quote}
$x<(U_{1},U_{2})$ if $x\in U_{1}$ (equivalently $x<U_{2}$),\\
$(U_{1},U_{2})<x$ if $x\in U_{2}$ (equivalently $x>U_{1}$),\\
$(U_{1},U_{2})<(U_{1}^{\prime},U_{2}^{\prime})$ if $U_{1}\subset U_{1}%
^{\prime}$ (equivalently $U_{2}^{\prime}\subset U_{2}$).
\end{quote}

We denote $U_{K}:=(U\uplus K,\leq)$.

\subsection{Functional trees}

Functional trees are the infinite terms written with function symbols of fixed
arity that are used in semantics of program (\cite{Cou83,CouHdbk9}) and also
(see below Section 2.3) for defining labelled linear orders called
\emph{arrangements}. These trees can be formalized in different ways.\ In
order to avoid useless technicalities, we limit definitions to function
symbols of arity at most 2, as we will only use this case. We call them
$F$-terms to stress their algebraic nature.

\bigskip
\noindent \textbf{Definitions 2.3} : \emph{Functional trees, equivalently, infinite
terms.}

\smallskip
(a) A \emph{binary tree-domain} is a set of words $D\subseteq\{1,2\}^{\ast}$
(called \emph{Dewey words}) that is \emph{prefix-closed}, \ which means that
$u\in D$ if $uv\in D,$ and is such that $u1\in D$ if $u2\in D$. We consider
$(D,\leq)$ as a rooted tree, where the \emph{ancestor relation} $\leq$
reverses the prefix order on words (we get $uv\leq u$).\ Its root is the empty
word $\varepsilon$.

\medskip
(b) Let $F$ be a finite \emph{binary functional signature},\emph{ i.e.}, a
finite set of symbols equipped with an \emph{arity mapping} $\rho
:F\rightarrow\{0,1,2\}$. A \emph{functional tree over} $F$, called an
$F$-\emph{term}\ for short is a pair $t=(D_{t},lab_{t})$ consisting of a
binary tree-domain $D_{t}$ and a mapping $lab_{t}: D_{t} \rightarrow F$ such that,
for every $u$ in $D_{t}$, $\rho(lab_{t}(u))$ is the number of integers
$i\in\{1,2\}$ such that $ui\in D_{t}$. (See Examples 2.5).\ The set of $F$-terms
is denoted by $T^{\infty}(F)$ and the subset of finite ones by $T(F)$.\ We
call $u\in D_{t}$ a \emph{position} in $t$ (or a \emph{node} if we think of
$t$ as a tree); it is an $\emph{occurrence}$ of the function symbol
$lab_{t}(u).$ We denote also by $Pos(t)$ the set of positions of $t$.

The ancestor relation on $Pos(t)$ is denoted by $\leq_{t}$.

\medskip
(c) If $t=(D_{t},lab_{t})$ and $u\in D_{t}$, then $t/u$ is the $F$-term with
set of positions $D_{t}/u:=\{w\in\{1,2\}^{\ast}\mid uw\in D_{t}\}$ and
labelling function such that $lab_{t/u}(w):=lab_{t}(uw)$. We call $t/u$ the
subterm \emph{issued from} $u$.\ If $u\neq v$\ and the $F$-terms $t/u$ and
$t/v$ are equal, then the corresponding subtrees\footnote{In the sense of
graph theory.} of the labelled rooted tree $t$ are isomorphic but not equal,
because their sets of nodes are $\{uw\in\{1,2\}^{\ast}\mid uw\in D_{t}\}$ and
$\{vw\in\{1,2\}^{\ast}\mid vw\in D_{t}\}$.

\medskip
(d) We recall that the \emph{lexicographic order} on $\{1,2\}^{\ast}$ is
defined as follows:

\begin{quote}
$u\leq_{lex}v$ if and only if $v=uv^{\prime}$ or $u=w1u^{\prime}$ and
$v=w2v^{\prime}$ for some words $w,u^{\prime}$ and $v^{\prime}$.
\end{quote}

It is a linear order. Hence, it defines a linear order on every tree-domain.
Another order will be defined in Definition 2.17. \ $\square$

\bigskip
 \noindent \textbf{Definitions 2.4} : \emph{Operations on }$F$-\emph{terms}.

(a) Let $f\in F$ be of arity 2.\ Let $t_{1}$ and $t_{2}\in T^{\infty}%
(F)$.\ \ Then $t:=f(t_{1},t_{2})$ is defined as follows:\smallskip
\begin{quote}
$D_{t}:=\{\varepsilon\}\uplus1D_{t_{1}}\uplus2D_{t_{2}},$

$lab_{t}(\varepsilon):=f,lab_{t}(1u):=lab_{t_{1}}(u),lab_{t}(2u):=lab_{t_{2}%
}(u).$
\end{quote}

If $f$ has arity 1 and $t_{1}\in T^{\infty}(F)$, then, $t:=f(t_{1})$ is
defined as follows:\smallskip
\begin{quote}
$D_{t}:=\{\varepsilon\}\uplus1D_{t_{1}},$

$lab_{t}(\varepsilon):=f$ and $lab_{t}(1u):=lab_{t_{1}}(u).$
\end{quote}

If $t=f$, a nullary symbol, then $D_{t}:=\{\varepsilon\}$ and $lab_{t}%
(\varepsilon):=f.\square$

\bigskip
The link between trees and terms is best illustrated by examples.

\bigskip
 \noindent\textbf{Examples 2.5} : Let $F:=\{f,g,a,b\}$ where these symbols have
respective arities 2,1,0 and 0.

(1) The term $t=f(g(a),f(a,b))$ has domain $D_{t}=\{\varepsilon
,1,2,11,21,22\}$\ and labelling function $lab_{t}$ such that : \smallskip
\begin{quote}
$\varepsilon\longmapsto f,1\longmapsto g,2\longmapsto f,11\longmapsto
a,21\longmapsto a,22\longmapsto b$.
\end{quote}

(2) The infinite term $s=f(a,f(g(b),f(a,f(g(b),....)))$ has domain
$D_{s}:=2^{\ast}\uplus(22)^{\ast}1\uplus2(22)^{\ast}1\uplus2(22)^{\ast}11$ and
labelling function $lab_{s}$ such that : \smallskip
\begin{quote}
$2^{i}\longmapsto f,(22)^{i}1\longmapsto a,2(22)^{i}1\longmapsto
g,2(22)^{i}11\longmapsto b$ for all $i\geq0$. \ $\square$
\end{quote}

%\bigskip
 \noindent\textbf{Definition 2.6} : \emph{Regular terms.}

An $F$-term $t$ is \emph{regular} if the set of its subtrees is finite. Every
finite $F$-term is regular.

\bigskip
Proposition 2.10\ states equivalent characterizations of regular terms, for
which we review some definitions.

\bigskip
 \noindent\textbf{Definition 2.7} : \emph{Automata.}

(a) A \emph{finite automaton} over $F$ (as in Definition 2.3(b)) is (in the
present article) a tuple $\mathcal{B}=(F,S,\tau,s_{init})$ where $S$ is the
finite set of \emph{states}, $s_{init}$ is the \emph{initial state} and $\tau$
is the \emph{transition function,} a total mapping : \smallskip

$S\rightarrow F\times((S\times S)\uplus S\uplus\{\varepsilon\})$ \smallskip

such that, for each state $s$, $\tau(s)$ is either $(f,s_{1},s_{2})$ if $f$
has arity 2, or $(f,s_{1})$ if $f$ has arity 1 or $(f,\varepsilon)$ if $f$ has
arity 0, for some $s_{1},s_{2}\in S$ and $f\in F$.\medskip

(b) Let $t\in T^{\infty}(F).$ \emph{The} \emph{run} of $\mathcal{B}$ on $t$ is
the (unique) mapping $run_{\mathcal{B}}:Pos(t)\rightarrow S$ such that
$run_{\mathcal{B}}(\varepsilon)=s_{init}$ and, for every $u\in Pos(t),$ if $u$
is an occurrence of $f$ and $run_{\mathcal{B}}(u)=s$, we have the following cases:
\begin{quote}
if $\rho(f)=2$, then $\tau(s)=(f,run_{\mathcal{B}}(u1),run_{\mathcal{B}}(u2))$,\\
if $\rho(f)=1$, then $\tau(s)=(f,run_{\mathcal{B}}(u1))$, and\\
if $\rho(f)=0$, then $\tau(s)=(f,\varepsilon)$.
\end{quote}

Hence, a state $s$ at a position $u$ defines (via $\tau$) the symbol at $u$
and the states at the positions $u1$ and $u2$ when they do exist. There is at
most one run \ $run_{\mathcal{B}}$.\ It is defined, deterministically, in a
top-down way.\medskip

(c) An $F$-term $t$ is \emph{accepted by} $\mathcal{B}$\ if $\mathcal{B}$ has
a run on $t$. As automata are top-down deterministic, each of them accepts
exactly one \emph{F}-term.

\bigskip
 \noindent\textbf{Definition 2.8 }: \emph{First-order (FO) and monadic second-order
(MSO)\ logic }

Let $\mathcal{R}$ be a finite set of relation symbols $\{R_{1},...,R_{k}\}$,
each being given with a positive integer $\rho(R_{i})$ called its
\emph{arity.}\medskip

(a) A $\mathcal{R}$-\emph{relational structure} is a tuple $S=(D_{S}%
,R_{1S},...,R_{kS})$ where $D_{S}$\ is the \emph{domain} and \ $R_{iS}$ is a
$\rho(R_{i})$-ary relation on the domain. Its properties will be expressed by
first-order (FO) or monadic second-order (MSO) formulas or sentences.\ A
\emph{sentence} is a formula without free variables.\medskip

(b) A partial order is by definition a relational structure. For\ expressing
properties of an $F$-term $t,$ we will use the relational structure: \smallskip
\begin{quote}
$\left\lfloor t\right\rfloor :=(D_{t},son_{1},son_{2},(lab_{f})_{f\in F})$
where $D_{t}$\ is the domain,\\
$son_{i}(u,v):\Longleftrightarrow$ $v=ui$,\\
$lab_{f}(u):\Longleftrightarrow$ $u$ is an occurrence of $f$.
\end{quote}\smallskip

(c) The finiteness of a set $X$ is not expressible in MSO\ logic. Hence, we
will use MSO formulas written with a finiteness set predicate $Fin(X)$, and
denote by MSO$_{fin}$ the corresponding extension of MSO logic. However, if
$X\subseteq V$, $V$ is partially ordered and $X$ linearly ordered, then its
finiteness is MSO-expressible in terms of the order relation of $V$
\cite{CouLMCS}.

\bigskip
 \noindent\textbf{Definition 2.9}: \emph{Regular equation systems}

Let $t_{1},...,t_{n}$ be unknowns denoting $F$-terms.\ A \emph{regular
equation system} is an $n$-tuple $(t_{i}=s_{i};i=1,..,n)$ where each $s_{i}$
is of the form $f(t_{j},t_{k})$ or $f(t_{j})$ or $f$, the symbol $f$ has arity
respectively 2,1 or 0, and $j,k\in\{1,..,n\}.$

A \emph{solution} is an $n$-tuple of $F$-terms that satisfy the equations.
Every such system has a unique solution (see \cite{Cou83}). $\square$

\bigskip
 \noindent\textbf{Proposition 2.10} : An $F$-term $t$ is regular if and only if
\begin{itemize}
\itemsep=-0.5pt
\item[(i)] it is a component of the unique solution of a regular equation system,
\item[(ii)] it is accepted by a finite automaton,
\item[(iii)] for each $f\in F$, the set of occurrences of $f$ in $t$ is a regular
language (included in $\{1,2\}^{\ast}$).
\item[(iv)] the structure$\ \left\lfloor t\right\rfloor $ is the unique model
\emph{u.t.i.} (that means \emph{up to isomorphism}) of an MSO sentence.
\end{itemize}

 \begin{proof}  Characterizations (i) and (ii) are clear from the
definitions.\ For (iii) see \cite{Cou83}.\ \ For (iv) see \cite{Tho90}%
\end{proof}%.$\square$
\eject

 \noindent\textbf{Examples 2.11} : (1) The $F$-term $s$ of Example 2.5(2) is regular, as
it is defined by the equation system:\smallskip
\begin{quote}
\{$s=f(r_{a},u),u=f(w,s),w=g(r_{b}),r_{a}=a,r_{b}=b\}$.
\end{quote}

The different subterms are $s,u,w,r_{a}$ and $r_{b}$. An automaton accepting
it has states $s,u,w,r_{a},r_{b}$, initial state $s$ and transitions
$s\longmapsto(f,r_{a},u),u\longmapsto(f,w,s),w\longmapsto(g,r_{b}%
),r_{a}\longmapsto(a,\varepsilon),r_{b}\longmapsto(b,\varepsilon)$.

Other examples will be given in Examples 2.16.\medskip

(2) Let $t$ be regular and accepted by an automaton $\mathcal{B}%
=(F,S,\tau,s_{init})$.\ We let $t_{\mathcal{B}}$\ be obtained by replacing a
symbol $f$ at position $u$ by the symbol $(f,run_{\mathcal{B}}(u))$.\ Then
$t_{\mathcal{B}}$ is a regular ($F\times S)$-term defined by an automaton
having the same states as $\mathcal{B}$. The arity of $(f,s)$ is that of $f$.
\ $\square$

\bigskip
 \noindent\textbf{Definition 2.12} : \emph{MSO definable sets of positions of an
F-term.}

Let $t$ be an $F$-term.\ Let $\varphi(X_{1},...,X_{n})$ be an MSO\ formula
such that there is a unique tuple $X_{1},...X_{n}$ of sets of positions of $t$
such that $\ \left\lfloor t\right\rfloor \models\varphi(X_{1},...,X_{n})$.\ We
say that $(X_{1},...,X_{n})$ is \emph{MSO-definable} in $t$.\

Let now $F^{\prime}:=F\times\mathcal{P}([n])$ and $t_{\varphi}$\ the
$F^{\prime}$-term obtained from $t$ by replacing a symbol $f$ at an occurrence
$u$ by the symbol $(f,I)$ of same arity where $I$ is the set of indices $i$
such that $u\in X_{i}$, where $(X_{1},...,X_{n})$ is defined by $\varphi$. We
have the following.

\bigskip
 \noindent\textbf{Lemma 2.13 } : Let $t$ be regular and $F^{\prime}$ and $\varphi$\ be
as above.

(1) The $F^{\prime}$-term $t_{\varphi}$\ is regular.

(2) In the case where $n=1$, the cardinality of the unique set $X_{1}$ such
that $\left\lfloor t\right\rfloor \models\varphi(X_{1})$ can be computed.

 \begin{proof} Easy consequences of Proposition 2.10 together with routine
logical manipulations.
\end{proof}%\ $\square$

\subsection{Arrangements and labelled sets}

We review a notion introduced in \cite{Cou78} and further studied in
\cite{Hei,Tho86}.\

\bigskip
 \noindent\textbf{Definitions 2.14} : \ \emph{Arrangements}

\medskip
(a) Let $X$ be a countable set. A linear order $(V,\leq)$ equipped with a
labelling mapping $lab:V\rightarrow X$ is called an \emph{arrangement}
\emph{over} $X$. We denote by $\mathcal{A}(X)$ the\ set of arrangements over
$X$. A linear order $(V,\leq)$ is identified with the arrangement
$(V,\leq,Id)$ such that $Id(v):=v$ for each $v\in V$.\

An arrangement over a finite set $X$\ can be considered as a generalized word
over alphabet $X$.

\medskip
(b) An \emph{isomorphism of arrangements} $i:(V,\leq,lab)\rightarrow
(V^{\prime},\leq^{\prime},lab^{\prime})$ is an order preserving bijection
$i:V\rightarrow V^{\prime}$ such that $lab^{\prime}\circ i=lab.$ Isomorphism
is denoted by $\simeq$ (as for all structures).

\medskip
(c) The concatenation of linear orders (denoted by the noncommutative
operation +) yields a concatenation of arrangements denoted by $\bullet$. We
denote by $\Omega$\ the empty arrangement and by $a$ the one reduced to a
single occurrence of $a\in X$. Clearly, $w\bullet\Omega=\Omega\bullet w=w$ for
every $w\in\mathcal{A}(X)$. \ The infinite word $w=a^{\omega}$ is the
arrangement over $\{a\}$ with underlying linear order $\omega$; it is
described by the equation $w=a\bullet w$.\ Similarly, the arrangement
$w=a^{\eta}$ \ over $\{a\}$ with underlying linear order $(\mathbb{Q},\leq)$
(that of rational numbers) is described by the equation $w=w\bullet(a\bullet
w)$.

\bigskip
 \noindent\textbf{Definition 2.15} : \emph{Terms defining arrangements.}

(a) Let $X$ be a set of nullary symbols and $t\in T^{\infty}(\{\bullet
,\Omega\}\uplus X).$ The set $Pos(t)$\ of positions of $t$ is the tree-domain
\ $D_{t}\subseteq\{1,2\}^{\ast}$. The \emph{value }of $t$ is the arrangement
$val(t):=(\mathrm{Occ}(t,X),\leq_{lex},lab)$\ where $\mathrm{Occ}(t,X)$ is the
set of positions of the elements of $X$ and $lab(u)$ is the symbol of $X$
occurring at position $u$.\ We say that $t$ \emph{denotes }an
arrangement\emph{ }$w$ if $w$ is isomorphic to $val(t)$.

\medskip
(b) An arrangement is \emph{regular} if it is denoted by a regular
term.\ $\square$

\bigskip
 \noindent\textbf{Examples 2.16 } : (a) $t_{0}:=\bullet(a,\bullet(b,\bullet(a,\bullet
(b,\bullet(.........))))))$\ denotes the infinite word $abab...$ .\ Its value
is defined from the set of words $\mathrm{Occ}(t_{0},\{a,b\})=2^{\ast}1$,
lexicographically ordered\footnote{We have $1<21<221<...$} and the labelling
function such that $lab(2^{i}1):=a$ if $i$ is even and $lab(2^{i}1):=b$
if\ $i$ is odd. The term $t_{0}$ is regular.\

\medskip
(b) The arrangements\ $a^{\omega}$ and $a^{\eta}$ (whose underlying orders
are, respectively, the natural and rational numbers) are denoted respectively
by $t_{1}$ and $t_{2}$ that are the unique solutions in $T^{\infty}%
(\{\bullet,\Omega,a\})$ of the equations $t_{1}=a\bullet t_{1}$ and
$t_{2}=t_{2}\bullet(a\bullet t_{2})$. These two arrangements are regular.

\smallskip
The term $t_{2}$ is defined from the two equations $t_{2}=t_{2}\bullet s$ and
$s=a\bullet t_{2}.$ The tree-domains of $t_{2}$ and $s$ satisfy the equalities :
\begin{quote}
$D_{t_{2}}=\varepsilon\cup1D_{t_{2}}\cup2D_{s}$ and $D_{s}=\varepsilon
\cup1\cup2D_{t_{2}},$
\end{quote}
\noindent hence $D_{t_{2}}$\ is defined by the regular expression $(1\cup22)^{\ast
}(\varepsilon\cup2\cup21).\ $The positions in $(1\cup22)^{\ast}21$ are the
occurrences of $a$. $\ \square$

\medskip
An arrangement is regular\footnote{Equivalently is a component of the
\emph{initial solution of a regular equation system }over $\{\bullet
,\Omega\}\uplus X$ ($X$ is finite, cf.\ \cite{Cou78}).\ We will not use this
characterization.} if and only if it is the value of a \emph{regular
expression} in the sense of \cite{Hei}. The later characterization implies
that regularity is preserved under alphabetic homomorphisms : if
$w=(V,\leq,lab)\in\mathcal{A}(X)$ is regular and $r:X\rightarrow Y$ is a
partial mapping, then $(V^{\prime},\leq,r\circ lab)$ is a regular arrangement
over $Y$ where $V^{\prime}$ is the set of $x\in V$ such that$\ r(lab(x))$ is
defined\footnote{A letter $a$ is erased if $r(a)$ is undefined.}.

We will also use the result of \cite{Tho86} that an arrangement over a finite
alphabet is regular if and only if is MSO-definable, where we represent an
arrangement $w=(V,\leq,lab)$ over $X$ by the relational structure
$\left\lfloor w\right\rfloor :=(V,\leq,(lab_{a})_{a\in X})$ such that
$lab_{a}(u)$ is true if and only if $lab(u)=a$.

\bigskip
 \noindent\textbf{Definition 2.17} : \emph{The inorder on words over }$\{1,2\}.$

We define as follows a strict linear order $<_{in}$ on $W:=\{1,2\}^{\ast}$:\smallskip
\begin{quote}
$x<_{in}y$ :$\Longleftrightarrow$\ either $x=y1x^{\prime}$, or $y=x2y^{\prime
},$ or $x=z1x^{\prime}$ and $y=z2y^{\prime}$ for some words $x^{\prime
},y^{\prime},z.$
\end{quote}

In the last two cases, we have $x<_{lex}y$. $\square$
\eject

%\bigskip
The verification that $<_{in}$ is indeed a strict linear order is easy from
definitions.\ It generalizes to infinite binary trees the \emph{inorder} on
the nodes of finite ones.

\bigskip
 \noindent\textbf{Proposition 2.18} : Let $F$ be a finite set of function symbols of
arity at most 2.\ Let $t\in T^{\infty}(F)$ be a regular term.\

(1) The arrangement $\widehat{t}:=(Pos(t),\leq_{in},lab_{t})$ over $F$ is regular.

(2) If $X\subseteq Pos(t)$ is MSO-definable in $t$, then the arrangement
$\widehat{t}[X]:=(X,$ $\leq_{in},lab_{t})$\ is regular.\medskip

\textbf{Proof sketch: }(1)\textbf{ }Let $t=t_{1}$ be defined by equations
$t_{i}=s_{i}$ where $i=1,..,n.$ We obtain as follows a finite equation systems
defining the arrangements $\widehat{t_{i}}$ for $i=1,..,n.$ \smallskip

If $t_{i}=f(t_{j},t_{k}$) then $\widehat{t_{i}}=(\widehat{t_{j}}\bullet
f\ )\bullet\widehat{t_{k}}.$

If $t_{i}=f(t_{j}$) then $\widehat{t_{i}}=\widehat{t_{j}}\bullet f.$

If $t_{i}=f$ then $\widehat{t_{i}}=f.$ \medskip

(2) By Lemma 2.13\ and\ the fact that the regularity of arrangements is
preserved by erasing letters in a deterministc way.\ $\square$

 \medskip
 \noindent\textbf{Definition 2.19} : \emph{Labelled sets, or commutative arrangements}

(a) An $X$-\emph{labelled set} is a pair $m=(V,lab)$ where $V$ is a set and
$lab$\ is a mapping $:V\rightarrow X$,\ equivalently, if $X$ is finite, a
relational structure $(V,(lab_{a})_{a\in X})$ where each element of $V$
belongs to a unique set $lab_{a}$.\ We denote by $\mathcal{S}(X)$ the set of
$X$-labelled sets. \medskip

(b) We denote by $set(w)$ the $X$-labelled set obtained by forgetting the
linear order of an arrangement $w$ over $X$.\ A term in $t\in T^{\infty
}(\{\bullet,\Omega\}\cup X)$ defines the $X$-labelled set $set(val(t))$. Over
labelled sets, the operation $\bullet$\ \ is commutative. \medskip

Up to isomorphism, an $X$-labelled set $m$ is defined by the cardinalities in
$\mathbb{N}\cup\{\omega\}$ of the sets $lab_{a}$, hence is a countable
\emph{multiset of elements of} $X$ : a number in $\mathbb{N}\cup\{\omega\}$ is
associated with each $a\in X$ and represents its number of occurrences in $m$.

If $X$ is finite, each $X$-labelled set is MSO$_{\mathit{fin}}$-definable,
\emph{i.e.}, is the unique countable model, \emph{u.t.i.}, of a sentence of
\emph{monadic second-order logic} extended with a set predicate $Fin(U)$
expressing that a set $U$ is finite. It is also \emph{regular},\emph{ i.e.},
is $set(val(t))$ for some regular term in $T^{\infty}(\{\bullet,\Omega\}\cup
X).$ The notion of regularity is thus trivial for $X$-labelled sets when $X$
is finite.

\section{Order-theoretic trees and forests}

In order to have a simple terminology, we will use the prefix\emph{ O-} to
mean \emph{order-theoretic} and to distinguish these generalized trees from
the ordered ones in \cite{CouLMCS}. An order-theoretic forest is called simply
a \emph{tree} by Fra\"{\i}ss\'{e} in \cite{Fra}. We will distinguish carefully
trees, forests, O-trees and O-forests.

\bigskip
 \noindent\textbf{Definition 3.1 }: \emph{Order-theoretic forests and trees}.

(a) An \emph{O-forest} is a pair $J=(N,\leq)$ such that:\smallskip
\begin{quote}
1) $N$\ is a countable\footnote{Countable means finite or countably infinite.}
set called the set of \emph{nodes},
\end{quote}
\begin{quote}
2) $\leq$ is a partial order on $N$\ such that, for every node $x,$ the set
$[x,+\infty\lbrack$ (the set of nodes $y\geq x$) is linearly ordered.
\end{quote}

It is called an \emph{O-tree} if furthermore: \smallskip
\begin{quote}
3) every two nodes $x$ and $y$ have an upper-bound.
\end{quote}

An O-forest $(N,\leq)$ is the disjoint union of O-trees whose sets of nodes
are the connected components of the comparability graph of $\leq$. More
precisely, two nodes are in a same component, \emph{i.e.} in the same
composing O-tree, if and only if they have a (common) upper bound. \medskip

(b) A minimal node is a \emph{leaf}.\ If $N$ has a largest element $r$ (that
is $x\leq r$ for all $x\in N$) then $J$\ is a \emph{rooted }O-tree and $r$ is
its \emph{root}.\ In an O-tree, the set of strict upper-bounds of a nonempty
set $X\subseteq N$ is an upwards closed line\footnote{See Section 2.1\ for the
notion of line.\ That a set $A$ is \emph{upwards closed} means that
$[u,+\infty\lbrack\subseteq A$ for all $u\in A$.} $L$.\  \medskip

(c) An O-tree is a \emph{join-tree}\footnote{An \emph{ordered tree} is a
rooted tree such that the set of sons of each $x$ is linearly ordered by an
order depending on $x$.\ This notion is extended in \cite{CouLMCS} \ to
join-trees.\ Ordered join-trees should not be confused with order-theoretic
trees.} if every two nodes $x$ and $y$ have a least upper-bound denoted by
$x\sqcup y$ and called their \emph{join} (cf.\ Section 1).\ It is a
\emph{join-forest} if every two nodes having an upper-bound actually have a
join. $\ \square$

\bigskip

If $T$ is a finite rooted tree, then $(N_{T},\leq_{T})$ is a join-tree
($\leq_{T}$ is the ancestor relation) and every finite O-tree is a join-tree
of this form.

\subsection{Structurings}

O-forests will be partitioned into lines, \emph{i.e.}, into convex linearly
ordered subsets.

\bigskip
 \noindent\textbf{Definition 3.2}\ : \emph{Covering between lines.}

Let $J=(N,\leq)$ be an O-forest. If $U$ and $W$ are two lines, we say that $W$
\emph{covers} $U$, denoted by $U\prec W$, if $U<w$ for some $w$ in $W$ and,
for all $x\in N$ and $w$ in $W$, if $U<x<w,$ then $x\in W$.\ Hence, there is
nothing between $U$ and $W$: if $U<y$ and $y\in N$, then there is $w\in W$
such that $U<w\leq y$. The covering relation is not transitive, hence $\prec$
is not a strict partial order.

\bigskip
 \noindent\textbf{Definitions 3.3 }: \emph{Structurings of O-forests }

(a) Let $J=(N,\leq)$ be an O-forest.\ A \emph{structuring} of $J$ is a set
$\mathcal{U}$ of nonempty lines that forms a partition of $N$ and satisfies
the following condition:

For each $x$ in $N$, we have\footnote{The set $[x,+\infty\lbrack$\ may have a
greatest element that this notation does not specify.} $[x,+\infty
\lbrack=I_{k}\uplus I_{k-1}\uplus...\uplus I_{0}$ for nonempty intervals
$I_{0},...,I_{k}$ of $[x,+\infty\lbrack$ such that:\smallskip
\begin{quote}
$I_{k}<I_{k-1}<...<I_{0}$,\\
for each $j$, we have $I_{j}\subseteq U$ for some unique line $U$ in
$\mathcal{U}$, denoted by $U_{j}$,\\
each interval $I_{j}$ is upwards closed in $(U_{j},\leq)$, which implies that
$U_{j}\neq U_{j^{\prime}}$ if $j\neq j^{\prime}.$
\end{quote}

It follows that $U_{k}\prec U_{k-1}\prec...\prec U_{0}$ and that $I_{0}$ and
$U_{0}$ are upwards closed in $J$.

Then $J=(N,\leq,\mathcal{U})$ is a \emph{structured O-forest}, an
\emph{SO-forest} in short.

A\emph{ }structuring of an O-forest is a union of structurings of the O-trees
composing it. \medskip

(b) The sequence $I_{0},I_{1},...,I_{k}$ is uniquely defined for each $x$, and
$k$ is called the \emph{depth }of $x$. If $x\in N$, then $U(x)$ denotes the
line of $\mathcal{U}$ containing $x.$ We define $\beta(x):=[x,+\infty
\lbrack-U(x)=I_{k-1}\uplus...\uplus I_{0}$ (hence, $x\notin\beta(x)$). \medskip

(c) If $J=(N,\leq,\mathcal{U})$ is an \emph{SO-forest} and $X\subseteq N$,
then the set of nonempty sets $X\cap U$ for $U\in\mathcal{U}$ is a structuring
of the O-forest $J[X]$.\ However, the depth of an element of $X$ may be
smaller in $J[X]$ than in $J$. $\square$

\bigskip
 \noindent\textbf{Example 3.4} : \emph{ }Figure 1 shows a structuring $\{A,B,C,D,E\}$ of
an O-tree. The line $A$ is upwards closed (we will call it the \emph{axis}%
).\ The depth of $x$ is 0, the depths of $y$ and $z$ are 2. We have
$\beta(x)=\emptyset$ \ and $\beta(u)=I$ where $u$ is any node in $B\cup D$.
The lines $\{A,B,C,D,E\}$ form a kind of tree. \ $\square$

\begin{figure}[!h]
\vspace*{2mm}
\begin{center}
\includegraphics[width=2.5244in]{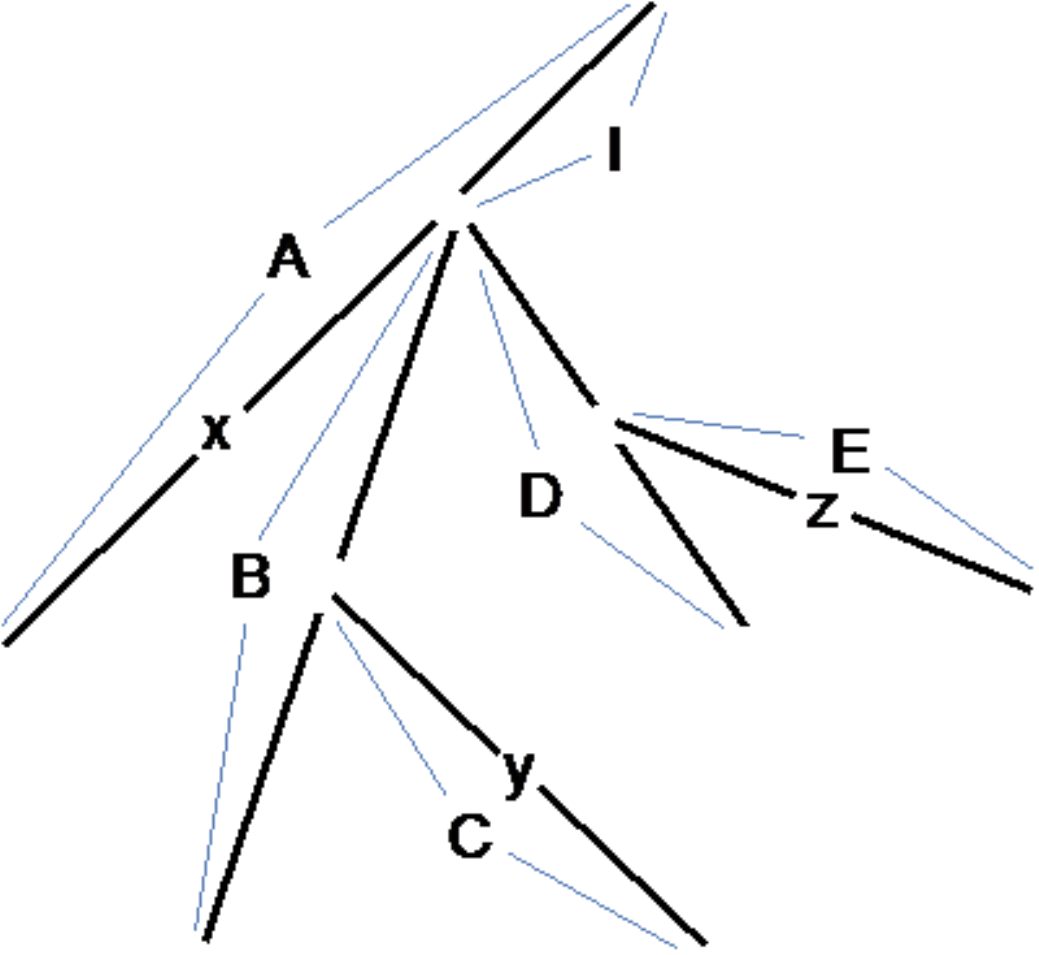}%
\caption{A structuring of an O-tree.}%
\end{center}\vspace*{-2mm}
\end{figure}
%EndExpansion

 \noindent\textbf{Proposition 3.5 }: Every O-forest has a structuring.

 %\medskip
\begin{proof} The proof is similar to that of \cite{CouLMCS}\ establishing
that every join-tree has a structuring. We give it for completeness. \medskip

We first consider an O-tree $J=(N,\leq)$.\ We choose an enumeration
$x_{0},x_{1},...,$\ $x_{n},...$\ of $N$ and a maximal\footnote{Maximal for set
inclusion.} line $B_{0}$; it is upwards closed.\ For each $i>0$, we choose a
maximal line $B_{i}$ containing the first node not in $B_{i-1}\cup...\cup
B_{0}$. We define $U_{0}:=B_{0}$ and, for $i>0$, $U_{i}:=B_{i}-(U_{i-1}%
\uplus...\uplus U_{0})$\ \ \ $=B_{i}-(B_{i-1}\cup...\cup B_{0})$. We define
$\mathcal{U}$ as the set of lines $U_{i}$. It is a structuring of $J$.  \medskip

An O-forest has a structuring that is the union of structurings of the O-trees
composing it. %$\ \ \square$
\end{proof}

\bigskip
 \noindent\textbf{Definition 3.6 }: \emph{Axis}

(a) An \emph{axis} of an SO-forest $J=(N,\leq,\mathcal{U})$ is a distinguished
upwards closed line $A\in\mathcal{U}$. It is an axis of one of the SO-trees
composing $J$. An \emph{SO-forest with axis}, an \emph{SOA-forest} in short,
is thus a 4-tuple $J=(N,\leq,\mathcal{U},A)$.\ If $A$ empty, we say that $J$
has no axis; it can be defined as $(N,\leq,\mathcal{U})$.

An nonempty SO-tree $J=(N,\leq,\mathcal{U})$ has a unique axis $A$, defined
from $\mathcal{U}$, that we denote by $Axis(J)$. Hence it is an SOA-tree in a
unique way.\ However, as for SO-forests, we may decide that it has no axis. \medskip

(b) The operation $\mathit{fg}$ \emph{forgets} the axis of an SOA-forest (or
of an SOA-tree):\smallskip
\begin{quote}
$\mathit{fg}(N,\leq,\mathcal{U},A):=(N,\leq,\mathcal{U})$ identified to
$(N,\leq,\mathcal{U},\emptyset)$.
\end{quote}\smallskip

(c) The union of pairwise disjoint SO-trees (without axes) is an SO-forest,
and conversely. If $J$ is an SO-forest, we denote by $Axes(J)$ the set of axes
of the SOA-trees composing it according to (a). \medskip

(d) If $J=(N,\leq,\mathcal{U})$ is an SO-forest and $U\in\mathcal{U}$, we
denote by $J\downarrow U$\ the SOA-tree $J[W]$ with axis $U$, where $W$ is the
union of the sets $]-\infty,x]$ for all $x\in U$.\ $\ \square$

\bigskip
 \noindent\textbf{Definition 3.7 :} \emph{Cuts defined from a structuring.}

Let $J=(N,\leq,\mathcal{U})$ be an SO-forest.

(a) If $U\in\mathcal{U}$, we say that a node $x\in N$ \emph{defines a cut}
$(U_{1},U_{2})$ of the linear order $(U,\leq)$ if $x\notin U$, $U_{2}%
=U\cap\lbrack x,+\infty\lbrack\neq\emptyset$ and $x\bot U_{1}$ (\emph{i.e.},
$x$ is incomparable with each element of the nonempty set $U_{1}$). This cut
is denoted by $\kappa(U,x)$. \medskip

(b) We let $Cuts(U)$ be the set of cuts $\kappa(U,x)$ of $U,$ and we denote by
$U_{Cuts(U)}$ the linearly orderered set $(U\uplus Cuts(U),\leq)$
(cf.\ Definition 2.2). It is countable as $N$\ is.  \medskip

(c) We denote by $\mathcal{K}$ the union of the sets $Cuts(U)$ for all
$U\in\mathcal{U}$. This set is countable because cuts are defined by nodes and
each node $x$ of depth $k>0$ defines finitely many cuts in lines of smaller depth.  \medskip

(d) If $\kappa$ is a cut of $U$ of depth $k\geq0,$ we denote by $\mathit{Def}%
(\kappa)$ the SO-forest (without axis) induced by the nodes $x$ such that
$\kappa(U,x)=\kappa.$ Then $Axes(\mathit{Def}(\kappa))$ is the set of lines
$W$ of $\mathit{Def}(\kappa)$ that are at depth $k+1$ (in $J$), hence such
that $W\prec U$.\ Each of them defines an SOA-tree $J\downarrow W$
and\ $\mathit{Def}(\kappa)$ is the union of the SO-trees $\mathit{fg}%
(J\downarrow W)$ for all $W\in Axes(\mathit{Def}(\kappa))$.  \medskip

We denote by $\mathcal{L}$ the set of O-forests $\mathit{Def}(\kappa)$ for all
$\kappa\in\mathcal{K}.$ It is in bijection with $\mathcal{K}$. \ \ $\square$

\bigskip
In Example 3.4\ and Figure 1, we let $\kappa:=\kappa(A,y)=\kappa(A,z).$ Then
$\mathit{Def}(\kappa)$ consists of the two trees $B\cup C$ and $D\cup E$,
$Axes(\mathit{Def}(\kappa))=\{B,D\}$ and $J\downarrow B$ is the SOA-tree
$B\cup C\ $ with axis $B$.

\bigskip
 \noindent\textbf{Lemma 3.8 }: Let $J=(N,\leq,\mathcal{U})$ be an SO-forest.

(1) Each node $x$ at depth $k>0$ defines a cut of some $U\in\mathcal{U}$ of
depth $k-1$.

(2) Each cut of $U$ of depth $k\geq0$ is $\kappa(U,x)$ for some node $x$ at
depth $k+1$.

 \medskip
  \noindent\textbf{Proof: }Clear from definitions. \ \ $\square$

\subsection{Description schemes of SOA-forests}

As observed above, a structuring of an O-tree can be seen as a kind of "tree
of lines".\ In a \emph{regular SO-tree} (a notion to be defined), the
structuring consists of finitely many lines \emph{up to isomorphism}, and they
are regular arrangements.\ By defining these arrangements by monadic
second-order sentences, we will obtain finitary descriptions.\ The formal
definitions are more involved.

\bigskip
 \noindent\textbf{Definitions and notation 3.9}\ : \emph{Concerning SO-forests}

Let $J=(N,\leq,\mathcal{U})$ be an SO-forest. It is a disjoint union of SO-trees. \medskip

(a) As observed in Definition 3.6(a), each of these SO-trees $L=(N_{L}%
,\leq,\mathcal{U}_{L})$\ has a unique axis $A\in\mathcal{U}_{L}\subseteq
\mathcal{U}$\ denoted by $Axis(L)$.\ We denote by $Axes(J)$ the set of lines
$Axis(L)$ for all these SO-trees $L$. \medskip

(b) Let $U\in\mathcal{U}.\ $Its \emph{tail}, denoted by $Tail(U),$ is the
SO-forest induced on $\{x\in N\mid x<U\}.\ $It may be empty\footnote{If we
define $\mathcal{U}$ by means of the construction of Proposition 3.5, all
tails are empty.}. We let $\mathcal{T}$ be the set of nonempty SO-forests
$Tail(U)$ for $U\in\mathcal{U}.\ $For each $U\in\mathcal{U}$, we let $\tau
_{U}$ be a new symbol, standing for $Tail(U)$ and located so as to mark that
$Tail(U)<U,$\ as shown by the next definition.\ Similarly, a cut $\kappa$ of
$U$ marks the position of $\mathit{Def}(\kappa)$ among the nodes of $U$. \medskip

(c) We recall (Definition 2.2) that $U_{Cuts(U)}$ is the arrangement extending
$U$ by inserting its cuts at their natual places. We let $U^{+}$ be the
arrangement $\tau_{U}\bullet U_{Cuts(U)}$ defined by prefixing $U_{Cuts(U)}$
with $\tau_{U}$. It is countable.

\bigskip
 \noindent\textbf{Definition 3.10} : \emph{Substitutions of SO-forests in linear
orders.}

Let $H=(N,\leq_{H})$ be a linear order, $X\subseteq N$ and for each $x\in X,$
let $F_{x}=(M_{x},\leq_{x},\mathcal{U}_{x})$ be an SO-forest such that the
sets $M_{x}$ are pairwise disjoint and disjoint from $N$. Then $H[x\leftarrow
F_{x};x\in X]$ is the SOA-forest $(M,\leq,\mathcal{U},A)$ such that : \smallskip
\begin{quote}
$M$ is the union of $N-X$ and the sets $M_{x}$,\\
$u\leq v$ if and only if \ $u,v\in N-X$ and $u\leq_{H}v$,\\
\qquad\qquad\qquad or\ $u,v\in M_{x}$ and $u\leq_{x}v$ for some $x\in X$,\ \\
\qquad\qquad\qquad or $v\in N-X$ and $u\in M_{x}$ for some $x\in X$ and
$x<_{H}v$. \smallskip \\
$A:=N-X$, \\
$\mathcal{U}$ consists of $A$ and the sets in the $\mathcal{U}_{x}$'s for all
$x\in X$. $\ \ \ \ \ \ \ \ \square$
\end{quote}

\smallskip
With these definition and notation, we have :

\bigskip
 \noindent\textbf{Lemma 3.11} : Let $J=(N,\leq,\mathcal{U})$ be an SO-forest and
$U\in\mathcal{U}$. Then we have:  \medskip

\qquad$J\downarrow U=U^{+}[\kappa\leftarrow\mathit{Def}(\kappa);\kappa\in
Cuts(U),\tau_{U}\leftarrow Tail(U)]$

if $Tail(U)$ is not empty.\ Otherwise,

$\qquad J\downarrow U=U^{+}[\kappa\leftarrow\mathit{Def}(\kappa);\kappa\in
Cuts(U)]$.

\begin{proof}
 Straightforward from the definitions. We have equalities, not
just isomorphisms. %$\ \square$
\end{proof}

\emph{Labellings of SO-forests.} \medskip

Let\emph{ }$J=(N,\leq,\mathcal{U})$ be an SO-forest, as in Definitions 3.7 and
3.9.\ Then $\mathcal{K}$ will denote the set of all cuts, $\mathcal{L}$ the
set of associated SO-forests $\mathit{Def}(\kappa)$ for all $\kappa
\in\mathcal{K}$ and $\mathcal{T}$ the set of all tails.\ Hence, $\mathcal{L}$
and $\mathcal{T}$ are disjoint sets of SO-forests.

 \medskip
Our objective is to use labelling functions $r:\mathcal{U}\rightarrow D$ and
$s:\mathcal{L\uplus T}\rightarrow Q,$ into disjoint sets $D$ and $Q,$ in such
a way that : \smallskip
\begin{quote}
$r(U)=r(U^{\prime})$ implies $J\downarrow U\simeq J\downarrow U^{\prime}$ and \\
$s(L)=s(L^{\prime})$ implies $L\simeq L^{\prime}.$
\end{quote}

 Actually, stronger conditions will be useful.

\bigskip
 \noindent\textbf{Definitions 3.12 }: \emph{Good labelling of an SO-forest J.}

A\emph{ good labelling} of $J=(N,\leq,\mathcal{U})$\ is defined from the
following items: \smallskip
\begin{quote}
disjoint sets $D$, $Q_{cut}$ and $Q_{tail}$, with $Q:=Q_{cut}\uplus Q_{tail}$, \\
mappings $r:\mathcal{U}\rightarrow D$ and $s:\mathcal{L\uplus T}\rightarrow Q$
such that: \\
\qquad$s(\mathcal{L)}\subseteq Q_{cut}$, $s(\mathcal{T)}\subseteq Q_{tail}$, \\
$\qquad r(U)=r(U^{\prime})$ implies $s\rhd U^{+}\simeq s\rhd U^{\prime+}$ and \\
$\qquad s(L)=s(L^{\prime})$ implies $r\{Axes(L)\}=r\{Axes(L^{\prime})\},$
\end{quote}

where we use the following notation:

\begin{quote}
$s\rhd U^{+}$ is the arrangement over $\{\ast\}\uplus Q,$ obtained by
replacing $\tau_{U}$ by $s(Tail(U))$, each $\kappa$ in $Cuts(U)$ by
$s(\mathit{Def}(\kappa))$ and each $u\in U$ by $\ast$,

$r\{Axes(L)\}$ is the multiset of elements of $D$ (cf.\ Definition 2.19)
consisting of the labels $r(Axis(W))$ for all $W$ in $Axes(L)$, hence
$r\{Axes(L)\}\in\mathcal{S}(D).$
\end{quote}

\smallskip
If $r$ and $s$ define a good labelling, then $r(U)=r(U^{\prime})$ implies
$s\rhd U^{+}\simeq s\rhd U^{\prime+}$ whence also $J\downarrow U\simeq
J\downarrow U^{\prime}$ (by Lemma 3.11), and $s(L)=s(L^{\prime})$ implies
$r\{Axes(L)\}=r\{Axes(L^{\prime})\}$ whence $L\simeq L^{\prime}.$

In what follows, we focus on O-trees. Extending the results to O-forests will
be staightforward as they are disjoint unions of O-trees.

\bigskip
 \noindent\textbf{Definition 3.13} : \emph{Description schemes.}

(a) A \emph{description scheme} is a tuple $\Delta=(D,Q,d_{Ax},(m_{q})_{q\in
Q},(w_{d})_{d\in D})$ where $D$ and $Q$ are disjoint sets, $d_{Ax}\in D,$
$m_{q}\in\mathcal{S}(D)$ for each $q\in Q$ and each $w_{d}$ is an arrangement
over $\{\ast\}\uplus Q$.  \medskip

(b) It describes an SOA-tree $J=(N,\leq,\mathcal{U},A)$\ if $J$ has a good
labelling with mappings $r:\mathcal{U}\rightarrow D$ and $s:\mathcal{L\uplus
T}\rightarrow Q$ such that :  \medskip

$r(A)=d_{Ax},$

if $r(U)=d$, then $w_{d}\simeq s\rhd U^{+}$,

if $s(L)=q$, then $m_{q}=r\{Axes(L)\}.$  \medskip

\noindent In that case, we say that $\Delta$ describes as well the SO-tree
$(N,\leq,\mathcal{U})$\ and the O-tree $(N,\leq)$. \medskip

(c) A description scheme $\Delta=(D,Q,d_{Ax},(m_{q})_{q\in Q},(w_{d})_{d\in
D})$ is \emph{regular} if $D\uplus Q$ is finite and each arrangement $w_{d}$
is regular.\ The finiteness of $D$ implies that the $D$-labelled sets $m_{q}$
are regular (cf. Section 2.3).

\bigskip
 \noindent\textbf{Lemma 3.14 } : (1) Every SOA-tree is described by some description scheme.

(2) Every description scheme describes a unique SOA-tree,\emph{ u.t.i.}

\begin{proof} (1) One can take $D:=\mathcal{U},Q:=\mathcal{L\uplus T}$ and
identity mappings $r$ and $s$.\

\noindent (2) Similar to the proof of Proposition 3.24\ in \cite{CouLMCS}.
\end{proof}

 \smallskip
 \noindent\textbf{Examples 3.15}\ : (1) If all components $D,Q$,$m_{q}$ and $w_{d}$ of a
description scheme $\Delta$ are finite, then, the defined SOA-tree is a
regular, possibly infinite, rooted tree (with an axis).

\medskip
For example, $\Delta:=(\{d\},\{q\},d,m_{q},w_{d})$ where $m_{q}=\{d\}$ and
$w_{d}=q\ast$ defines the SOA-tree $J:=(\mathbb{N},\leq_{J},\mathcal{U}%
,\{0\})$ such that $n\leq_{J}m$ if and only if $m\leq n$ and $\mathcal{U}$ is
the set of singletons $\{n\}$.

\begin{figure}[!h]
\vspace*{1mm}
\begin{center}
\includegraphics[width=3in]{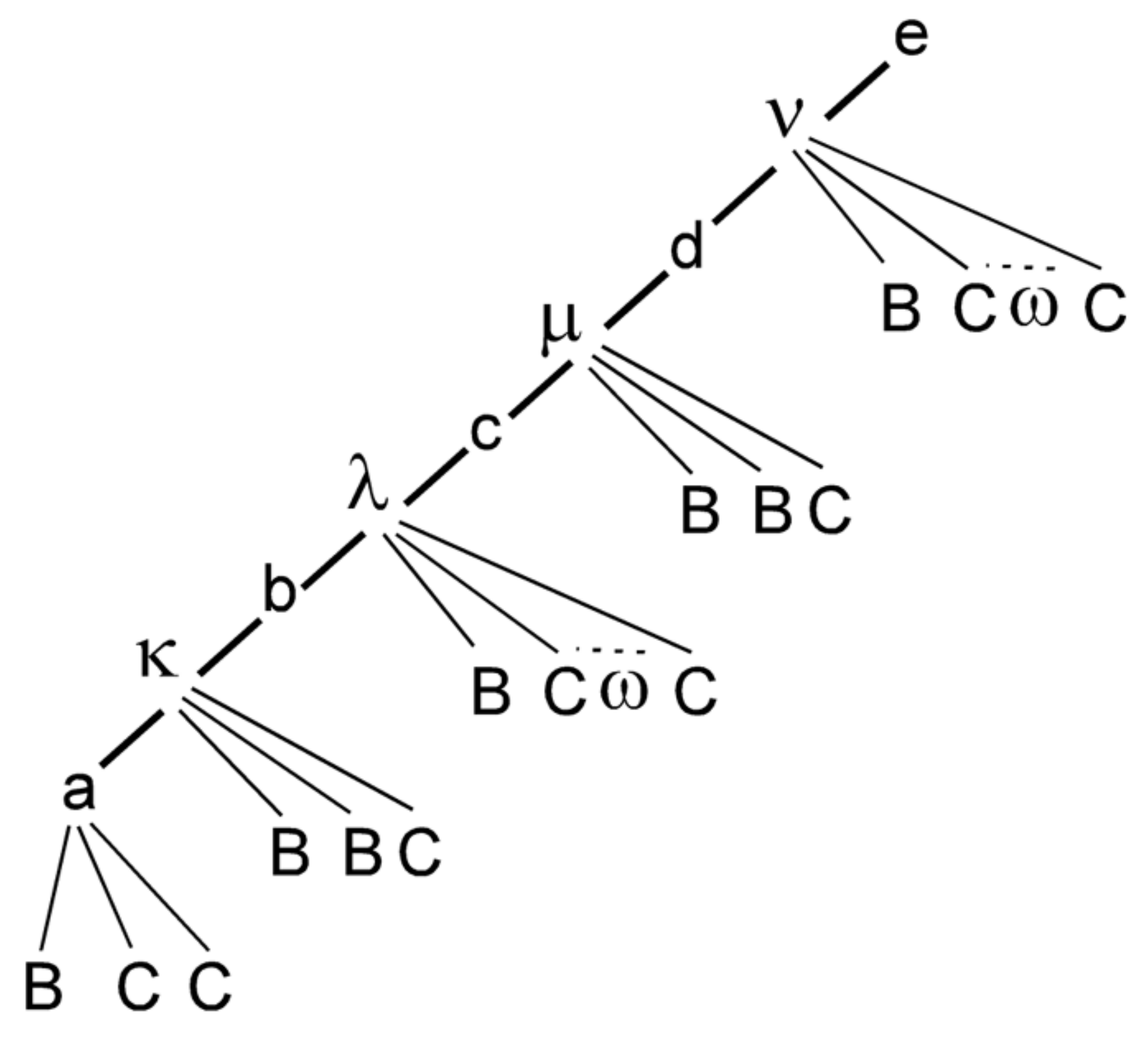}\vspace*{-1.45cm}
\caption{See Example 3.15.}%
\end{center}\vspace*{-1mm}
\end{figure}

(2) Figure 2 sketches a description scheme of a structured O-tree whose lines
have depth 0 or 1. The axis $A$ is $\{a,b,c,d,e\}$, it has four cuts named
$\kappa,\lambda,\mu,\nu.$ The other lines of the structuring have isomorphism
types $B$ or $C$.\ The SO-forests $\mathit{Def}(\kappa)$ and $\mathit{Def}%
(\mu)$ are both isomorphic to the multiset $\{B,B,C\}$: we mean that the
SO-forest $\mathit{Def}(\kappa)$ has three components respectively isomorphic
to $B,B$ and $C$.\ Similarly, $\mathit{Def}(\lambda)$ and $\mathit{Def}(\nu)$
are isomorphic to $\{B,C,...,C\}$ with $\omega$ times $C$. The axis has a tail
isomorphic to $\{B,C,C\}.$ We let $p$ label $\kappa$ and $\mu$, $q$ label
$\lambda$ and $\nu$, and $z$ label $\tau_{A}$. Then $s\rhd A^{+}=z\ast p\ast
q\ast p\ast q\ast$ (cf.\ Definition 3.12).

A corresponding description scheme has $D:=\{\alpha,\beta,\gamma\}$,
$Q:=\{z,p,q\}$, $d_{Ax}:=\alpha$, $w_{\alpha}:=s\rhd A^{+},w_{\beta}:=s\rhd
B,w_{\gamma}:=s\rhd C$, $m_{z}:=\{\beta,\gamma,\gamma\},$ $m_{p}%
:=\{\beta,\beta,\gamma\},$ $m_{q}:=\{\beta,\gamma,...,\gamma\}$ with $\omega$
times $\gamma$. As $B$ and $C$\ have neither cuts nor tails, $B^{+}=B,C^{+}=C$
and the arrangements $s\rhd B$ and $s\rhd C$ are over $\{\ast\}$.

The elements of $D$ name the isomorphism types of the lines that form the
structuring.\ The elements of $Q$ name the isomorphism types of the axes of
the SOA-trees composing the SO-forests defined by the tail of $A$ and by its
cuts. More than the isomorphism type of the axis $A$, the element $\alpha$ of
$D$ designates the arrangement $s\rhd A^{+}$ that incorporates descriptions of
the cuts of $A$ (and the SO-forests they define) and of its tail. \ $\square$%

\subsection{Monadic second-order description of SO-forests}

In view of our use of monadic second-order logic, we give a description of
SO-forests by relational structures. An SO-forest $J=(N,\leq,\mathcal{U})$ is
not \emph{per se} a relational structure\ because $\mathcal{U}$ is a partition
of $N$\ in an unbounded number of sets. We will encode $\mathcal{U}$ by a
bipartition of $N$, by using a tool introduced in Definition 3.6\ of
\cite{CouLMCS}.

\bigskip
 \noindent\textbf{Definition 3.16} : \emph{SO-forests represented by relational
structures.}

If $J=(N,\leq,\mathcal{U})$ is an SO-forest, we define $S(J)$ as the
relational structure $(N,\leq,N_{0},N_{1})$\ such that $N_{0}$ is the set of
nodes at even depth and $N_{1}:=N-N_{0}$ is the set of those at odd depth;
$N_{0}$ and $N_{1}$ are sets but we consider them also as unary relations.

\bigskip
 \noindent\textbf{Proposition 3.17} :(1) There is an MSO\ formula $\varphi(N_{0}%
,N_{1})$\ expressing that a relational structure $(N,\leq,N_{0},N_{1})$\ is
$S(J)$ for some SO-forest $J=(N,\leq,\mathcal{U})$.\medskip

(2) There exist MSO\ formulas $\theta_{1}(N_{0},N_{1},U)$\ and $\theta
_{2}(N_{0},N_{1},U,W)$\ expressing respectively, in a structure $(N,\leq
,N_{0},N_{1})=S((N,\leq,\mathcal{U})),$ the properties "$U\in\mathcal{U}$" and
"$U,W\in\mathcal{U}$ and $U\prec W$" (cf.\ Definition 3.2 for $\prec$\ ).

\begin{proof}
 Easy extension of Proposition 3.7 of \cite{CouLMCS}%
\end{proof}

We recall from Definition 3.6 that if $(N,\leq,N_{0},N_{1})$\ is $S(J)$ for
some SO-tree $J=(N,\leq,\mathcal{U})$, then its only possible axis is the
unique set $A$ in $\mathcal{U}$ such that $A\subseteq N_{0}$ and there is no
$x\in N_{1}$ such that $A<x$.

\medskip
One of our key results is the following theorem.

\bigskip
 \noindent \textbf{Theorem 3.18} : Let $\Delta$ be a regular description scheme.\ There
exists an MSO$_{fin}$ sentence that characterizes \emph{u.t.i.} the SOA-trees
described by $\Delta$.

\medskip
Its proof will be given after some more notation and definitions.

\bigskip
 \noindent\textbf{Definitions 3.19} : \emph{Defining nodes}.\

Let $J=(N,\leq,\mathcal{U},A)$ be an SOA-tree. We use the notation of
Definitions 3.7, 3.9, 3.12 and 3.13. \medskip

(a) We say that $x\in N$ \emph{defines\footnote{A different notion of
representation of a line by an element (actually a position in a
term\ defining $J$) will be used in Sec\-tion~4.}} $W\in\mathcal{U}$ if
\ $W=U(x)$. \medskip

We say that $x\in N$ \emph{defines the tail of} $W\in\mathcal{U}$ if $U(x)<W$
and $U(x)\prec W$, which implies that $Tail(W)$ is not empty and
$U(x)\subseteq Tail(W)$. Furthermore, $U(x)\in Axes(Tail(W))$.

We say that $x\in N$ \emph{defines the cut} $(W_{1},W_{2})=\kappa$\emph{ }of
$W\in\mathcal{U}$ if $U(x)<W_{2}$, $U(x)\bot W_{1}$ and $U(x)\prec W$, which
implies that $U(x)\subseteq\mathit{Def}(\kappa)$ and $U(x)\in
Axes(\mathit{Def}(\kappa))$ (cf.\ Definition 3.9(b)). \medskip

(b) A triple of sets $(N_{\mathcal{U}},N_{\mathcal{T}},N_{\mathcal{K}})$ is
\emph{well-defining} (for $J$) if :
\begin{quote}
$N_{\mathcal{U}}$\ contains exactly one element defining each $W\in
\mathcal{U}$, \smallskip \\
$N_{\mathcal{T}}$ contains exactly one element defining a nonempty tail
$Tail(W)$ for all $W\in\mathcal{U}$,  \smallskip \\
$N_{\mathcal{K}}$ contains exactly one element defining each cut in
$\mathcal{K}$, the set of all cuts of all lines.
\end{quote}

We have $N_{\mathcal{T}}\cap N_{\mathcal{K}}=\varnothing.$\medskip

(c) Let $p,p^{\prime}$ be positive integers.\ A mapping $r:\mathcal{U}%
\rightarrow\lbrack p]$ is described by a partition $(R_{1},...,R_{p})$ of
$N_{\mathcal{U}}$, such that $r(U)=i$ if and only if there is some $x\in U\cap
N_{\mathcal{U}}\cap R_{i}$. A mapping $s:\mathcal{T\uplus L}\rightarrow\lbrack
p^{\prime}]$ is described similarly by a partition $(S_{1},...,S_{p^{\prime}%
})$ of $N_{\mathcal{T}}\uplus N_{\mathcal{K}}$. (We recall that $\mathit{Def}$
defines a bijection $\mathcal{K}\rightarrow\mathcal{L}$).\medskip

(d) Let $N_{0},N_{1}$\ be as in Definition 3.16.\ A pair $(r,s)$ that defines
a good labelling (cf.  Definition 3.12) where $D=[p]$ and $Q=[p^{\prime}]$ can
be described by a tuple of sets $(N_{\mathcal{U}},N_{\mathcal{T}%
},N_{\mathcal{K}},R_{1},...,R_{p},S_{1},...$, $S_{p^{\prime}})$\ that satisfies
$\varphi(N_{0},N_{1})$\ and, thanks to Proposition 3.17(2), the conditions of
(b) and (c).\medskip

We denote by $\psi(N_{0},N_{1},N_{\mathcal{U}},N_{\mathcal{T}},N_{\mathcal{K}%
},R_{1},...,R_{p},S_{1},...,S_{p^{\prime}})$ the MSO-formula expressing the
conjunction of these conditions (including $\ \varphi(N_{0},N_{1})).$
$\ \ \square$

\medskip \noindent
\textbf{Proof of Theorem 3.18} : \ Let $\Delta=(Q,D,d_{Ax},(m_{q})_{q\in
Q},(w_{d})_{d\in D})$ be a regular description scheme.

For each $d\in D$, there is an MSO-formula $\psi_{d}$ that characterizes,
\emph{u.t.i.}, the regular arrangement $w_{d}$, cf.\ Section 2.3.\ It is
written with the relation symbol $\leq$\ and unary relation symbols $lab_{q}$,
$lab_{\ast}$ used for encoding the labelling in $Q\uplus\{\ast\}$.

For each $q\in Q$, there is an MSO$_{fin}$-formula $\eta_{q}$ that
characterizes, \emph{u.t.i.}, the labelled set (the multiset) $m_{q}$.\ It is
written with the unary symbols $lab_{d}$ for encoding the labelling in $D$.
The finiteness predicates are necessary to distinguish the infinite sets from
the finite ones.\medskip

Without loss of generality, we assume that $D=[p]$ and $Q=[p^{\prime}]$.

\ Let $(N,\leq)$ be an O-tree. (An FO sentence can check that $(N,\leq)$ is
actually an O-tree). Let $N_{0},N_{1},N_{\mathcal{U}},N_{\mathcal{T}%
},N_{\mathcal{K}},R_{1},...,R_{p},S_{1},...,S_{p^{\prime}}$\ be $5+p+p^{\prime
}$ sets of nodes.\ The MSO\ formula $\psi(N_{0},N_{1}$, $N_{\mathcal{U}%
},N_{\mathcal{T}},N_{\mathcal{K}},R_{1},...,R_{p}$,$S_{1},...,S_{p^{\prime}})$
expresses that they define a structuring $\mathcal{U}$ of the O-tree
$(N,\leq)$, that the sets $N_{\mathcal{U}},N_{\mathcal{T}},N_{\mathcal{K}}$
and the mappings $r$ and $s$ on the sets $\mathcal{U}$ and $\mathcal{T}%
\cup\mathcal{K}$ are as in Definition 3.19(b,c).\

In order to build an MSO$_{fin}$ formula that checks the conditions of
Definition 3.13(b), we use the follows steps.\medskip

(a) For each $x\in R_{d}$,\ $1\leq d\leq p$, the arrangement $s\rhd U(x)^{+}$
has a domain consisting of $U(x)$ whose elements are labelled by $\ast$,
together with the nodes $y$ of $N_{\mathcal{K}}$\ that define cuts\footnote{We
replace in the definition of $U^{+}$ (Definition 3.9(c)) a cut by the node in
$N_{\mathcal{K}}$ that defines it.} of $U(x)$; each such $y$ is labelled by
$s(\mathit{Def}(V,W))\in Q_{cut}$ where $(V,W)$ is the cut it defines.
Furthermore, if the tail of $U(x)$ is not empty, then $s\rhd U(x)^{+}$ has a
first element $z$ in $N_{\mathcal{T}}$ marking its place, and that is labelled
by $s(Tail(U(x)))$. The linear ordering of $U(x)^{+}$ following from
Definition 2.2 is MSO-definable, as for each $y\in N_{\mathcal{K}}$, the
corresponding pair $(V,W)$ can be MSO-defined.

It remains to express that $s\rhd U(x)^{+}$ satisfies $\psi_{r(x)}$ which can
be done by an MSO\ formula, constructed from $\psi_{r(x)}$ by
\emph{relativization} to the domain of $U(x)^{+}$ described above.

In this way, we express that the arrangements $U^{+}$ associated with the
lines $U$\ of $\mathcal{U}$ satisfy Definition 3.13(b).\medskip

(b) Next we consider the cuts, defined by nodes $y\in S_{q}\cap N_{\mathcal{K}%
}.$ Let $y$ define a cut $\kappa$.\ We recall that $Axes(\mathit{Def}%
(\kappa))$ is the set of axes of the SOA-trees that form the SO-forest
$\mathit{Def}(\kappa)$.\ Each of these axes $A$ has a defining node
$\partial(A)$ in $N_{\mathcal{U}}$, and each $\partial(A)$ has an image in $D$
by $r$ (specified by ($R_{1},...,R_{p})$).\ We obtain a labelled set
$r\{Axes(\mathit{Def}(\kappa))\}$. By means of the MSO$_{fin}$ formula
$\eta_{s(y)}$, one can express that it is isomorphic to $m_{q}.$

The case of tails, defined by the nodes in the sets $S_{q}\cap N_{\mathcal{T}%
}$ can be treated similarly.\medskip

(c) We also express that the axis of the structuring of $(N,\leq)$ defined by
$N_{0},N_{1}$ contains a node in $R_{r(d_{Ax})}.$

\medskip
Hence, we get an MSO$_{fin}$ formula $\Theta_{\Delta}(N_{0},N_{1}%
,N_{\mathcal{U}},N_{\mathcal{T}},N_{\mathcal{K}},R_{1},...,R_{p},S_{1}%
,$\ \ $...,S_{p^{\prime}})$\ (it implies $\psi(N_{0},N_{1},N_{\mathcal{U}%
},N_{\mathcal{T}},N_{\mathcal{K}},R_{1},...,R_{p}$,$S_{1},...,S_{p^{\prime}}%
)$), that expresses the conditions of Definition 3.13(b) and 3.19(b,c).

\medskip
It follows that an O-tree $(N,\leq)$ has a structuring defined by $\Delta$ if
and only if it satisfies the MSO$_{fin}$ sentence:
\begin{quote}
$\exists N_{0},N_{1},N_{\mathcal{U}},N_{\mathcal{T}},N_{\mathcal{K}}%
,R_{1},...,R_{p},S_{1},...,S_{p^{\prime}}.\Theta_{\Delta}(N_{0}%
,...,S_{p^{\prime}})$ $.$ $\ \ \ \square$
\end{quote}

Our aim is now to establish the converse to Theorem 3.18 yielding Theorem
4.19\ that is our main theorem. We will use for that an algebra of SO-forests
of possible independent interest.

\section{The algebra of SO-forests}

We will actually use SO-forests enriched with distinguished axes. Their
algebra is similar to that of \cite{CouLMCS}\ for structured join-trees,\ but
it uses less operations and a single sort instead of two.

\bigskip
 \noindent\textbf{Definition 4.1 : }\emph{The algebra of SOA-forests.}

SOA-forests are defined in Definition 3.6. We define the following operations.\medskip

(a) \emph{Concatenation of SOA-forests along axes.}

\eject
The \emph{concatenation} $J\bullet J^{\prime}$ of disjoint SOA-forests
$J=(N,\leq,\mathcal{U},A)$ and $J^{\prime}=(N^{\prime},\leq^{\prime
},\mathcal{U}^{\prime},A^{\prime})$ is defined as follows:
\begin{quote}
$J\bullet J^{\prime}:=(N\uplus N^{\prime},\leq^{\prime\prime},\mathcal{U}%
^{\prime\prime},A\uplus A^{\prime})$ where :\\
$x\leq^{\prime\prime}y:\Longleftrightarrow x\leq y\vee x\leq^{\prime}%
y\vee(x\in N\wedge y\in A^{\prime}),$ \\
$\mathcal{U}^{\prime\prime}:=(\mathcal{U}-\{A\})\uplus(\mathcal{U}^{\prime
}-\{A^{\prime}\})\uplus\{A\uplus A^{\prime}\}$.
\end{quote}

This operation is associative.\ If $A=A^{\prime}=\emptyset,$ then $J\bullet
J^{\prime}$ is the union of $J$ and $J^{\prime}$, and then $J\bullet
J^{\prime}=J^{\prime}\bullet J.$

If $J$ and $J^{\prime}$ are not disjoint, we replace one of them by an
isomorphic copy disjoint from the other. The result is well-defined
\emph{u.t.i.} (we recall \emph{up to isomorphism}) as different isomorphic
copies can be chosen. (This is a standard technique.\ See \cite{CouEng}).

If $J$ and $J^{\prime}$ are linear orders with $A=N$, $\mathcal{U}%
=\{A\},A^{\prime}=N^{\prime}$ and $\mathcal{U}^{\prime}=\{A^{\prime}\}$ then
$J\bullet J^{\prime}$\ is their concatenation.

\begin{figure}[!h]
\vspace*{2mm}
\begin{center}
\includegraphics[width=4.7677in]{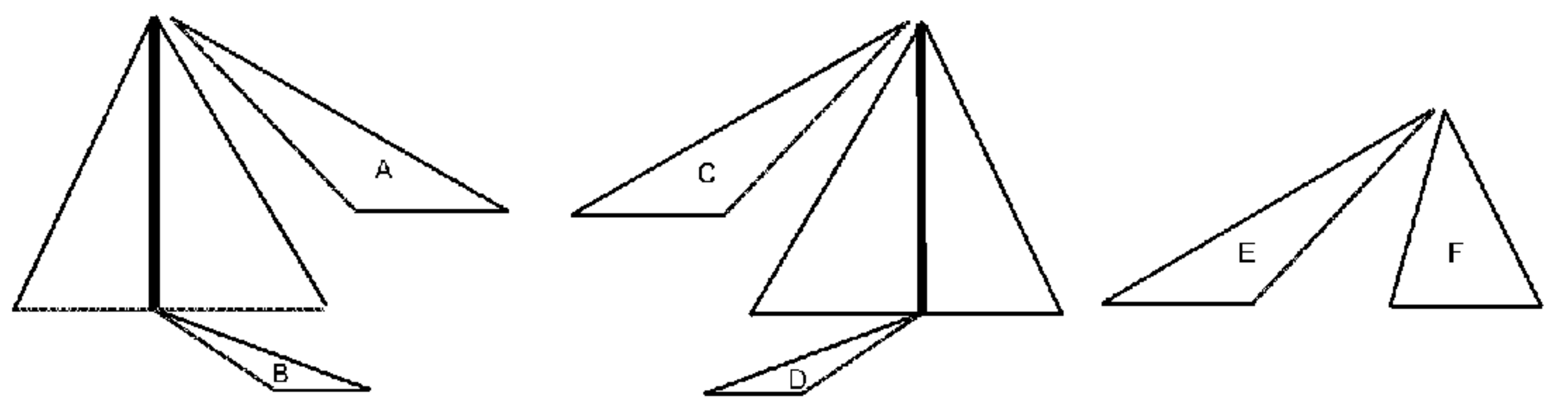}%
\caption{From left to right O-forests $J,J^{\prime}$ and $K$. Thick lines show
the axes. See Figure 4 for concatenations.}%
\end{center}
%\end{figure}
\vspace*{2mm}
%\begin{figure}[!h]
\begin{center}
\includegraphics[width=4.6484in]{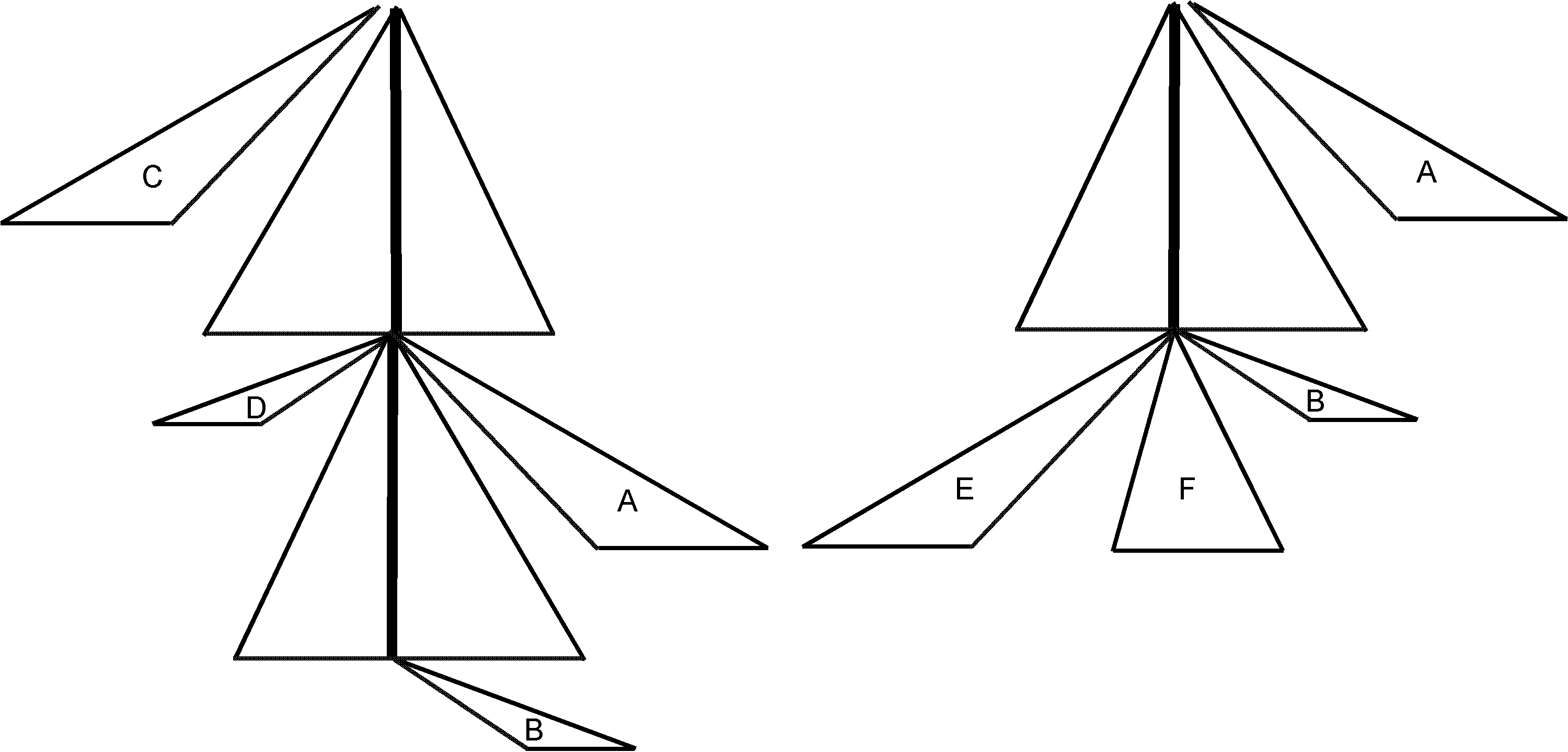}%
\caption{The concatenations $J\bullet J^{\prime}$\ and $K\bullet J.$}%
\end{center}\vspace*{-2mm}
\end{figure}
%EndExpansion

%\bigskip
Figure 3 shows SOA-forests $J,J^{\prime}$ and $K$ where $J$ and $J^{\prime}$
have axes shown by thick lines and $K$ has no axis.\ They all consist of two
SO-trees.\ Figure 4 shows the concatenations $J\bullet J^{\prime}$ and
$K\bullet J.$ Both consist of two O-trees.%

\medskip
(b) \emph{The empty SOA-forest }is denoted by the nullary symbol $\Omega$.

Clearly, $J\bullet\Omega=\Omega\bullet J=J.$

\medskip
(c) \emph{Nullary symbols for nodes}.\ For each $u$ (intended to be a node),
we denote by \underline{$u$} the SOA-tree $(\{u\},=,\{\{u\}\},\{u\}).$ When
considering SOA-trees and SOA-forests \emph{u.t.i.}, we replace every
\underline{$u$} by the unique symbol $\ast$ denoting \emph{any} singleton SOA-tree.

\medskip
(d) \emph{Forgetting the axis (cf.\ Definition 3.6).}

To forget the axis, equivalently, to make it empty, we use the unary operation
$\mathit{fg}$ such that\footnote{We could define an algebra with two sorts for
O-forests with and without an axis, and more operations.} :
\begin{quote}
$\mathit{fg}(N,\leq,\mathcal{U},A):=(N,\leq,\mathcal{U},\emptyset),$
equivalently, $(N,\leq,\mathcal{U}).$
\end{quote}

It is clear that:
\begin{quote}
$\mathit{fg}(\Omega)=\Omega,$  \smallskip\\
$\mathit{fg}(\mathit{fg}(J))=\mathit{fg}(J)$ and \smallskip \\
$\mathit{fg}(J)\bullet\mathit{fg}(J^{\prime})=\mathit{fg}(J^{\prime}%
)\bullet\mathit{fg}(J)=\mathit{fg}(J\bullet\mathit{fg}(J^{\prime})),$
\end{quote}

where $J$ and $J^{\prime}$ are disjoint or replaced by disjoint copies; in the
latter case, equality is replaced by isomorphism.

\medskip
(e) We let $F:=\{\bullet,\mathit{fg},\Omega,\ast\}$ and we obtain\ an
$F$-algebra of SOA-forests \emph{u.t.i,} denoted by $\mathbb{S}$. $\square$

\medskip
The value in $\mathbb{S}$ of a term in $T(F)$, \emph{i.e.}, of a finite term
over $F,$\ follows immediately from the definition of the operations. It is
defined \emph{u.t.i.} because we need to take disjoint copies of the
SO-forests that are arguments of concatenation: the simplest example is for
the term $\ast\bullet\ast.$\ Alternatively, \underline{$u$}$\bullet$%
\underline{$v$} denotes the SOA-tree $(\{u,v\},\leq,\{\{u,v\}\},\{u,v\})$
where $u<v$.

\begin{figure}[h]
\begin{center}
\includegraphics[width=3.8147in]{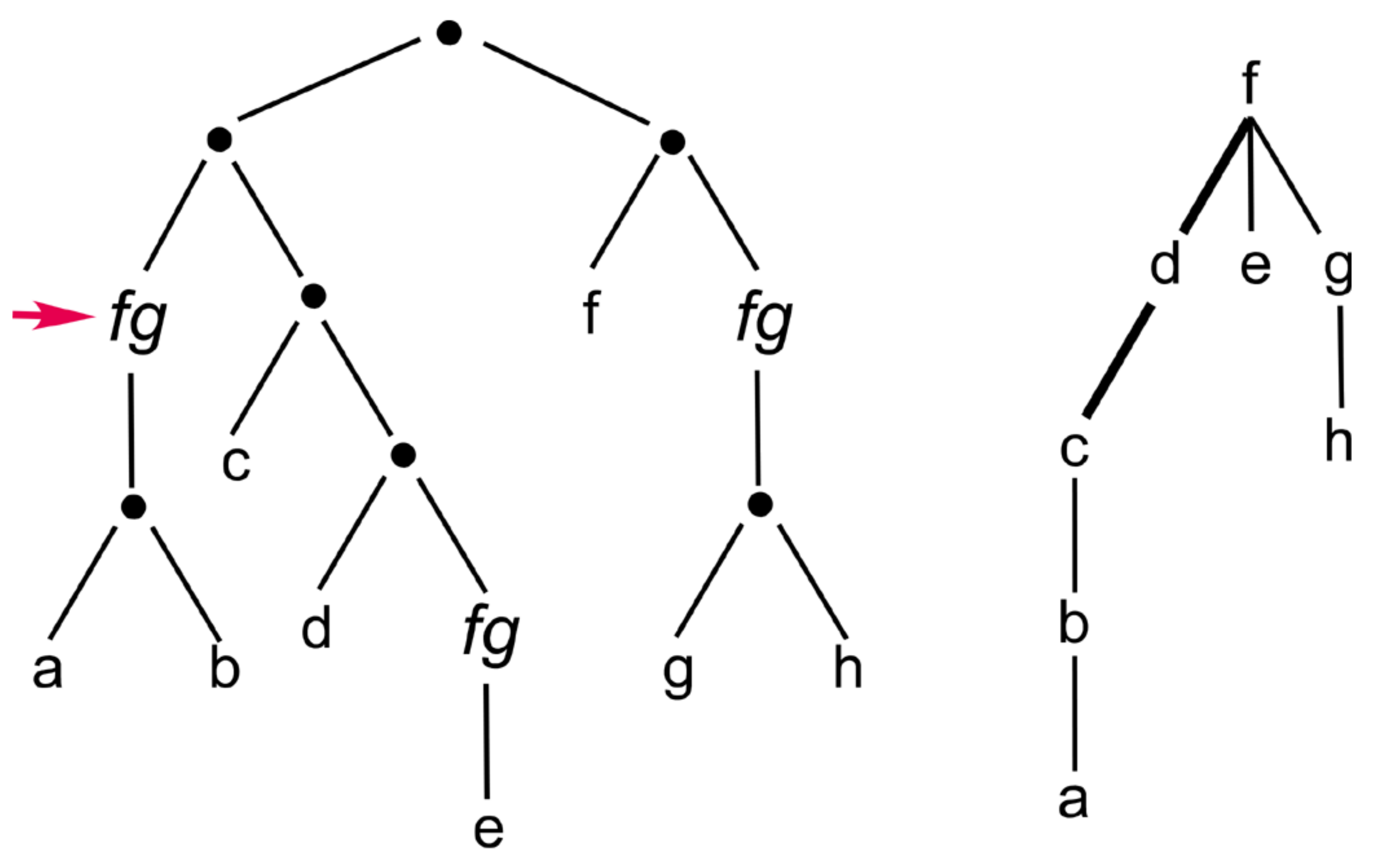}%
\caption{A term and the SOA-tree it defines, cf. Example 3.2.}%
\end{center}
\end{figure}
%EndExpansion

 \noindent
\textbf{Example 4.2}\ : Figure 5 shows a finite term $t$ (on the left) and the
SOA-tree $J$\ it defines.\ The bold edges indicate the axis $\{c,d,f\}.$ We
have $\mathcal{U}=\{\{a,b\},\{c,d,f\},$ $\{e\},\{g,h\}\}.$\ If in this term we
omit the operation $\mathit{fg}$ at position 11\ (we use Dewey notation for
positions, cf.\ Definition 2.3),\ marked in Figure 5 by an arrow, we obtain an
SOA-tree with axis $\{a,b,c,d,f\}$ and same ancestor relation.\ $\square$

\bigskip

Before giving a formal definition of the SOA-forest denoted by an infinite
term $t$, \emph{i.e.,} that will be its value in $\mathbb{S}$, we give three examples.

\bigskip
 \noindent
\textbf{Examples 4.3 } : \emph{Some SOA-forests and terms that denote them.}

(1) The SOA-tree $(\mathbb{Q},\leq,\{\mathbb{Q}\},\mathbb{Q})$ that we will
denote by $\mathbb{Q}$ is the value of the regular term $t_{0}$ defined as the
(unique) solution of the equation $t_{0}=t_{0}\bullet(\ast\bullet t_{0})$, cf.
Example 2.16(b) in Section 2.3\ about arrangements.\medskip

(2) Let now $J:=(\mathbb{Q},\leq^{\prime},\mathcal{U},\mathbb{Q}-B)$ be the
SOA-tree such that:
\begin{quote}
$B:=\{-n\mid n\in\mathbb{N}\}$, \smallskip\\
$x\leq^{\prime}y$ if and only if $x\leq y$\ \ and $y\notin B$, \smallskip\\
and its structuring $\mathcal{U}$ consists of the axis $\mathbb{Q}-B$ and the
sets $\{b\}$ for $b\in B$.
\end{quote}

It is not a join-tree because $(-n)\sqcup x$ is undefined if $n\in\mathbb{N}$
and $x<-n$.\ We have $J\simeq J\bullet(\mathit{fg}(\ast)\bullet\mathbb{Q}).$
It is the value of the term $t_{1}$ that is the unique solution of the
equation $t_{1}=t_{1}\bullet(\mathit{fg}(\ast)\bullet t_{0}).$ \medskip

(3) Let $H$ be an SOA-forest and $J:=\mathit{fg}(H)\bullet$\underline{$u$}.
Then $J$ is an SOA-tree with root $u$ and axis $\{u\}$; its subtrees below $u$
are the components of the forest $H$ whose axis has been "forgotten" by the
operation $\mathit{fg}$. This construction is denoted by $ext_{u}(H)$ in
\cite{CouLMCS}. \ \ $\square$

\medskip
We will use concrete terms that define SOA-forests with explicit nodes, as
opposed to SOA-forests up to isomorphism. For this purpose, if $M$\ is a set
of potential nodes, then \underline{$M$}\ is the set of nullary symbols
$\underline{u}$ for $u\in M$. We let $F_{\underline{M}}$ be the set
$\{\bullet,\mathit{fg},\Omega\}\uplus\underline{M}$. A \emph{concrete term} is
a term in\emph{ }$T^{\infty}(F_{\underline{M}})$ such that each \underline
{$u$}\ has at most one occurrence (because the arguments of $\bullet$\ must
define disjoint forests).\ Its value is an SOA-forest with nodes in $M$. The
following definition is similar to Definition 3.28 of~\cite{CouLMCS}.

\bigskip
 \noindent
\textbf{Definition 4.4 }: \emph{The value of a term in }$T^{\infty
}(F_{\underline{M}})$ \emph{or in} $T^{\infty}(F).$

Positions of terms are defined as words over $\{1,2\},$ cf.\ Definition 2.3. \medskip

(a) If $u,v$ are positions of a term $t$ and $v$ is an ancestor of $u$, we
write $u<_{t1}v$ if $u=v1u^{\prime}$ for some word $u^{\prime}$ (\emph{i.e.},
$u$ is the left son of $v$ or is below it), and similarly, $u<_{t2}v$ if
$u=v2u^{\prime}$. We will also use the lexicographic order $\leq_{lex}$ on words.

\medskip
(b) Let\ $t$ be a term in $T^{\infty}(F_{\underline{M}})$ or in $T^{\infty
}(F),$ and $w,w^{\prime}\in Pos(t).\ $We write $w\approx w^{\prime}$\ if and
only if $w=w^{\prime}$ or $w\neq w^{\prime}$ and there is no occurrence of the
operation $\mathit{fg}$ on the path in $t$ (considered as a rooted tree)
between its nodes $w$ and $w^{\prime}$.\ However, $w$ and/or $w^{\prime}$ may
be occurrences of $\mathit{fg}$ (cf.\ Example 4.10(2) below).

If $w$ and $w^{\prime}$ are occurrences respectively of \underline{$u$} and
\underline{$u^{\prime}$}, we write $u\approx u^{\prime}$ if and only if
$w\approx w^{\prime}$. We denote by $\mathrm{occ}(t,u)$ the occurrence of
$\underline{u}$ in $t$. We will frequently replace $\mathrm{occ}(t,u)$ by $u$.

\medskip
(c) \emph{The value of a concrete term. }Let $t\in T^{\infty}(F_{\underline
{M}}).$ We let $N$ be the set of $u\in M$ such that \underline{$u$} has one
occurrence in $t$ \ (hence a unique one). It is empty if, for example, $t$ is
$\Omega$ or $\Omega\bullet\Omega$.

\medskip
The \emph{value} $val(t)$ is the tuple $(N,\leq,\mathcal{U},A)$ defined as follows:\medskip

(i) Let $u,v\in N$.\ We define:
\begin{quote}
$u<v$ if and only if $\mathrm{occ}(t,u)<_{lex}\mathrm{occ}(t,v)$ and
$\mathrm{occ}(t,v)\approx w$ where $w:=\mathrm{occ}(t,u)\sqcup_{t}%
\mathrm{occ}(t,v)$ (the join of $\mathrm{occ}(t,u)$ and $\mathrm{occ}(t,v)$ in
$t$).
\end{quote}

Hence $w$ is an occurrence of $\bullet$, $\mathrm{occ}(t,u)<_{t1}w$ and
$\mathrm{occ}(t,v)<_{t2}w$. \medskip

(ii) The axis $A$\ is the set of nodes $u\in N$ such that $\mathrm{occ}%
(t,u)\approx root_{t}$ provided $root_{t}$ is not an occurrence of
$\mathit{fg}$; otherwise, there is no axis. \medskip

(iii) The sets in $\mathcal{U}$ are the nonempty sets $[x]_{\approx}\cap N$
for some position $x$ in $t$; each such set is $U(u)$ for some $u$ in $N$. On
a set $U(u),$ the order $\leq$ is the lexicographic order.\medskip

We will see that $val(t)$ is an SOA-forest.

\medskip
(d) \emph{The value of a term in} $T^{\infty}(F)$.\medskip

We construct from $t\in T^{\infty}(F)$ an SOA-forest $J=(N,\leq,\mathcal{U}%
,A)$ whose isomorphism class is the value of $t$, also denoted by $val(t)$:
\begin{quote}
$N:=\mathrm{Occ}(t,\ast)$, the set of occurrences of the nullary symbol $\ast$; \smallskip\\
the partial order $\leq$, the set $\mathcal{U}$ and the axis $A$ are defined
as above in (c).\
\end{quote}

For comparing (c) and (d), we observe that if $t^{\prime}\in T^{\infty
}(F_{\underline{M}})$ and $t\in T^{\infty}(F)$ is obtained from $t^{\prime}$
by substituting $\ast$ for each \underline{$u$}, then $val(t^{\prime})\simeq
val(t).$

\medskip
(e) An SOA-forest $J$ is \emph{regular} if it is, up to isomorphism, the value
of a regular term in $T^{\infty}(F)$. The SO-forest and the O-forest
underlying $J$ are said to be \emph{regular}.\ We say that a term $t^{\prime
}\in T^{\infty}(F_{\underline{M}})$ is regular if $t$ defined as in (d)
is.\ $\ \square$

\medskip
The logical structure $\left\lfloor t\right\rfloor $ representing a term $t$
is defined in Definition 2.8.

\bigskip
 \noindent
\textbf{Proposition 4.5} : If\ $t\in T^{\infty}(F_{\underline{M}}),$ then
$val(t)=(N,\leq,\mathcal{U},A)$ is an SOA-forest. The order $\leq$\ and the
set $\mathcal{U}$\ are MSO-definable in $\left\lfloor t\right\rfloor $. \medskip

\noindent \textbf{Proof sketch}: The following claims are easily proved from the
definitions and yield the stated facts.\ The pair $(N,\leq)$ is an O-forest.
The sets in $\mathcal{U}$ are linearly ordered and form a partition of $N$. If
$u<v<w$ and $u\approx w,$ then $u\approx v\approx w$: it follows that
$(N,\leq,\mathcal{U})$ is an SO-forest. Finally, $A\in\mathcal{U}$ and is an
axis. Definition 4.4(b) yields\ a monadic second-order definition in
$\left\lfloor t\right\rfloor $ of the structuring of $val(t)$ because the
equivalence relation $\approx$ is MSO definable. $\ \square$

\bigskip
 \noindent
\textbf{Proposition 4.6}\ : (1) If $t\in T^{\infty}(F_{\underline{M}})$,
$val(t)=(N,\leq,\mathcal{U},A)$\ and $X\subseteq N$, then the induced
SOA-forest $J[X]$ (cf. Definition 3.3(c)) is defined by the term $t^{\prime}$
obtained from $t$ by replacing, for each $u\notin X,$ the nullary symbol
$\underline{u}$\ by $\Omega$\ . \medskip

(2) Let $J_{i}$, $i=1,...$ be countably many pairwise disjoint SOA-forests,
defined respectively by concrete\ terms $t_{1},t_{2},...$.\ The term    $\mathit{fg}(t_{1})\bullet(\mathit{fg}(t_{2})\bullet...(\mathit{fg}%
(t_{i})\bullet(...)))$ defines an SO-forest without axis that is their union.
\eject

 \begin{proof}
  Clear from the definitions.
  \end{proof}

The following proposition justifies Definition 4.4.

 \medskip \noindent
\textbf{Proposition 4.7} : Let $t\bullet t^{\prime}\in T^{\infty
}(F_{\underline{M}})$.\ Then $val(t\bullet t^{\prime})=val(t)\bullet
val(t^{\prime}).$ We also have $val(\mathit{fg}(t))=\mathit{fg}(val(t))$,
$val(\Omega)=\Omega$ and $val(\underline{u})=u$.

 \begin{proof}
 Routine proofs from definitions.
\end{proof}

Hence, for a finite term $t$, the value $val(t)$ from Definition 4.4 is the
same as the one defined by induction on its structure. We also get that every
finite SOA-forest with set of nodes $N$\ is the value of a term in
$T(F_{\underline{N}})$.

\medskip \noindent
\textbf{Proposition 4.8} : Every SOA-forest $J$ is the value of a term in
$T^{\infty}(F_{\underline{N}})$ where $N$\ is its set of nodes.

 \begin{proof}
  Similar to Proposition 3.19\ in \cite{CouLMCS}.
 \end{proof}

\emph{Representations of lines, tails and cuts by positions.}

\medskip
In the following Lemmas 4.9, 4.11, 4.13 and 4.14, a term $t\in T^{\infty
}(F_{\underline{M}})$ defines a nonempty SOA-tree $J=(N,\leq,\mathcal{U},A).$
The ancestor relation in $t$ is denoted by $\leq_{t}$. As positions in $t$ are
words, we will also use on $Pos(t)$ the linear orders $\leq_{lex}$\ and
$\leq_{in}$. Furthermore, we will identify a node $u$ in $N$\ with
$\mathrm{occ}(t,u),$ the position in $t$ that defines it by being the unique
occurrence of $\underline{u}$. Hence, if $u,v\ \in N$, the notations $u\approx
v,$ $[u]_{\approx}$, $u\leq_{t}v,u\sqcup_{t}v,u\leq_{lex}v$ and $u\leq_{in}v$
will stand respectively for $\mathrm{occ}(t,u)\approx\mathrm{occ}(t,v),$
$[\mathrm{occ}(t,u)]_{\approx},$ $\ \mathrm{occ}(t,u)\leq_{t}\mathrm{occ}%
(t,v),\mathrm{occ}(t,u)\sqcup_{t}\mathrm{occ}(t,v),\mathrm{occ}(t,u)\leq
_{lex}\mathrm{occ}(t,v)$ and $\mathrm{occ}(t,u)\leq_{in}\mathrm{occ}(t,u).$
The notation $u\leq v$ refers to the order of $J$\ defined from $t$: here, $u$
and $v$ are nodes.

\medskip
 \noindent\textbf{Lemma 4.9 } : (1) For each $U=U(u)$ where $u\in N$, the set of
positions $[u]_{\approx}$ has a $\leq_{t}$-maximal element $x$ that is an
occurrence of $\bullet$ or of $\underline{u}$.\ We denote it by $Rep(U).$ \medskip

(2) For all $U,U^{\prime}$ in $\mathcal{U}$, if $t/Rep(U)\simeq
t/Rep(U^{\prime}),$ then $(U,\leq)\simeq(U^{\prime},\leq)$ and $Tail(U)\simeq
Tail(U^{\prime}).\square$

\medskip
Because of the second assertion, we say that position $Rep(U)$\ is
\emph{representative} of $U$ and, also, of $Tail(U)$. This notion differs from
that of a defining node in Definition 3.19.

 \begin{proof}
  (1) Consider $u\in N$ and $U:=U(u)$.\ The set $[u]_{\approx}$
has a unique maximal element $x$ with respect to $\leq_{t}.$ It cannot have
two because any two nodes have a join that is on the path linking them.

This position $x$ is an occurrence of a nullary symbol $\underline{w}$ for
some $w$ in $N$\ if and only if $w=u$ and $U=\{u\}.$ It can also be an
occurrence of $\bullet$.\ In these two cases, we define $Rep(U):=x$.

Otherwise, $x$ is an occurrence of $\mathit{fg}$ whose unique son $w$
\emph{must be} an occurrence of $\bullet$ or of $\underline{u}$ (because if
$w$ is also an occurrence of $\mathit{fg}$, then $x$ is not $\approx
$-equivalent to $u$).\ Then we define \mbox{$Rep(U):=w$}. %\medskip

\eject

(2) The occurrences of the nodes in $U$ are below $Rep(U)$ or equal to it.
Furthermore, the order $\leq$\ (of $val(t)$) restricted to $U$ is $\leq_{lex}%
$.\ It follows from Definition 4.4(a,b) that $(U,\leq)$ is fully defined from
$t/Rep(U)$ . Hence $(U,\leq)\simeq(U^{\prime},\leq)$ if $t/Rep(U)\simeq
t/Rep(U^{\prime})$.\medskip

Note that $t$ is a concrete term with nullary symbols denoting the nodes of
$val(t)$.\ Hence, our hypothesis is $t/Rep(U)\simeq t/Rep(U^{\prime})$.\ If
$t^{\prime}$ is obtained from $t$ by replacing each $\underline{u}$ by $\ast$
and $t^{\prime}/Rep(U)=t^{\prime}/Rep(U^{\prime})$, then we have also $(U,\leq)\simeq(U^{\prime},\leq)$.\
Similarly, all positions of $t$ that define $Tail(U)$ (we mean those that
define the nodes of $Tail(U)$ and those that are on the paths in $t$ between
any two nodes) are below $Rep(U)$ as one can check easily. (See Example
4.10(1)).\ Hence $Tail(U)\simeq Tail(U^{\prime})$ if $t/Rep(U)\simeq
t/Rep(U^{\prime})$.
\end{proof}

If $val(t)$ has a (nonempty) axis $A$, then $Rep(A)$ is the root of $t$. \

\bigskip
 \noindent\textbf{Examples 4.10}: (1) Consider the term $t_{0}:=\mathit{fg}%
(\underline{w})\bullet(\underline{u}\bullet\underline{v})$.\ The set
$U:=\{u,v\}$ is the axis of $val(t_{0})$ and $Rep(U)$\ is the root of $t_{0}$
(the position of the first occurrence of $\bullet$).\ We are in the third case
of the proof of Lemma 4.9(1). Note that $u\sqcup_{t}v$ (the position of the
second occurrence of $\bullet$), that is the join of (the positions of) two
elements of $U,$ is not the maximal element of $[u]_{\approx}$.\

The tail of $U$\ is $\{w\}$ (precisely $(\{w\},=,\{\{w\}\})$).\ The maximal
element of $[w]_{\approx}$ is the position of the unique occurrence of
$\mathit{fg}$; let $x$ be its son.\ We have $Rep(\{w\})=x$. This tail is
defined by $\mathit{fg}(\underline{w}),$ below $Rep(U)$.\medskip

(2) The term $t_{1}=\mathit{fg}(\Omega\bullet(\underline{a}\bullet
(\underline{b}\bullet\mathit{fg}(\Omega))))$ defines $J_{1}:=(\{a,b\},$%
\ $\leq,\{\{a,b\}\},\emptyset)$\ where $a<b$. The position of the first
occurrence of $\bullet$\ in $t$ is $Rep(\{a,b\}).\ $The two occurrences of
$\mathit{fg}$ are equivalent by $\approx$.\medskip

(3) Let $J_{2}:=(\{a,b,c,d,e,f\},$\ $\leq
,\{\{a,b\},\{c,d\},\{e\},\{f\}\},\{a,b\})$\ where $<$\ is defined from
$a<b,e<d<c<b$ and $f<d$ by transitivity. It is the value of the term
$\ \underline{a}\bullet\lbrack\mathit{fg}(\underline{d}\bullet\lbrack
\mathit{fg}(\underline{e})\bullet\mathit{fg}(\underline{f})]\ \underline
{\bullet}\ \underline{c})\bullet\underline{b}].$ The underlined occurrence of
$\bullet$, call\ it $x$, is representative of the line $\{c,d\}$ of
$\mathcal{U}$; the tail of this line is $(\{e,f\},=,\{\{e\},\{f\}\})$\ defined
by the subterm $\mathit{fg}(\underline{e})\bullet\mathit{fg}(\underline{f})$
that is actually below $x$. $\square$

\medskip
We now define representative positions of cuts. If $U\in\mathcal{U}$ and
$\kappa(U,x)=(U_{1},U_{2}),$ we recall that $U_{1}\bot x$ and $x<U_{2}.$ The
same cut is defined by each $z\in N$\ such that $U_{1}\bot z$ and $z<U_{2}.$

\bigskip
 \noindent\textbf{Lemma 4.11 } : Let $\kappa(U,x)=(U_{1},U_{2})$ be a cut of
$U.$\ Let\ $K$ be the set of joins $u\sqcup_{t}v$ for all $u\in U_{1}$ and
$v\in U_{2}$.\medskip

(1) If $u\in U_{1}$ and $v\in U_{2}$, we have $x<_{t}u\sqcup_{t}v$ . The set
$K$ is linearly ordered with respect to $<_{t}$.\ It has a minimal element
that we denote by $Rep(\kappa)$ and is an occurrence of $\bullet$. This
position is $\approx$-equivalent to $u$ and $v$.\medskip

(2) The subterm $t/Rep(\kappa)$ of $t$ defines an SOA-tree whose axis is the
interval of $(U,\leq_{lex})$ consisting of the elements $w$ of $U$ such that
$w<_{t}Rep(\kappa)$.\medskip

(3) The SO forest $\mathit{Def}(\kappa)$ is $val(t/Rep(\kappa))[L]$ where $L$
is the set of $x\in N$ such that $x<_{t}Rep(\kappa)$, $U_{1}\bot x$ and
$x<U_{2}.$\medskip

(4) If $\kappa$ is a cut of $U$ and $\kappa^{\prime}$ is a cut of $U^{\prime}$
such that\ $t/Rep(\kappa)\simeq t/Rep(\kappa^{\prime}),$ then $\mathit{Def}%
(\kappa)\simeq\mathit{Def}(\kappa^{\prime})$.

 \begin{proof} (1) Let $u\in U_{1}$ and $v\in U_{2}$.\ The position
$u\sqcup_{t}v$ is an occurrence of $\bullet$. As $u,v\in U,$ we have $u\approx
v$, and so, $u\approx u\sqcup_{t}v$.\medskip

We have $u<v$ and $x<v$, hence, by Definition 4.4, we have $u<_{t1}u\sqcup
_{t}v$ and $v<_{t2}u\sqcup_{t}v$, and, similarly $x<_{t1}x\sqcup_{t}v$ and
$v<_{t1}x\sqcup_{t}v$.\ Hence $u\sqcup_{t}v$ and $x\sqcup_{t}v$ are comparable
(with respect to $\leq_{t}).$

If $u\sqcup_{t}v<_{t}x\sqcup_{t}v$, then $x<_{t1}x\sqcup_{t}v$, $u<_{t}%
u\sqcup_{t}v<_{t2}x\sqcup_{t}v$.\ We also have $x\sqcup_{t}v\approx v,$ hence
$x\sqcup_{t}v\approx u\sqcup_{t}v$ as we have $u\approx u\sqcup_{t}v\approx
v.$ Hence, we have $x<u$, because $x<_{t1}x\sqcup_{t}v=x\sqcup_{t}u$,
$u<_{t2}x\sqcup_{t}u$ and $u\approx x\sqcup_{t}u.$ This contradicts the
assumption that $U_{1}\bot x.$ Hence $u\sqcup_{t}v\geq_{t}x\sqcup_{t}v$ and
$u\sqcup_{t}v>_{t}x.\ $

It follows that the set $K$ is linearly ordered in $t$ with respect to the
ancestor relation.\ It has a minimal element, say $w$, that we denote by
$Rep(\kappa)$. We have $\{u,x,v\}<_{t}Rep(\kappa)$ and $Rep(\kappa)\approx y$
for each $y\in U$.\medskip

(2) We also have $Rep(\kappa)\approx y$ for each $y$ in $[u]_{\approx}$ such
that $u\in U$.\ The axis of $val(t/Rep(\kappa))$ consists of the elements $u$
of $U$ that are below $Rep(\kappa)$ in $t$.\medskip

(3) and (4) are clear.\ Some leaves of $t/Rep(\kappa)$ not in $U$ may not
belong to $\mathit{Def}(\kappa)$ and so, may not define the cut $\kappa$.\ See
Example 4.12.
\end{proof}

We will say that the position $Rep(\kappa)$ is \emph{representative} of the
cut $\kappa$.

\begin{figure}[!b]
\begin{center}
\includegraphics[width=3.0372in]{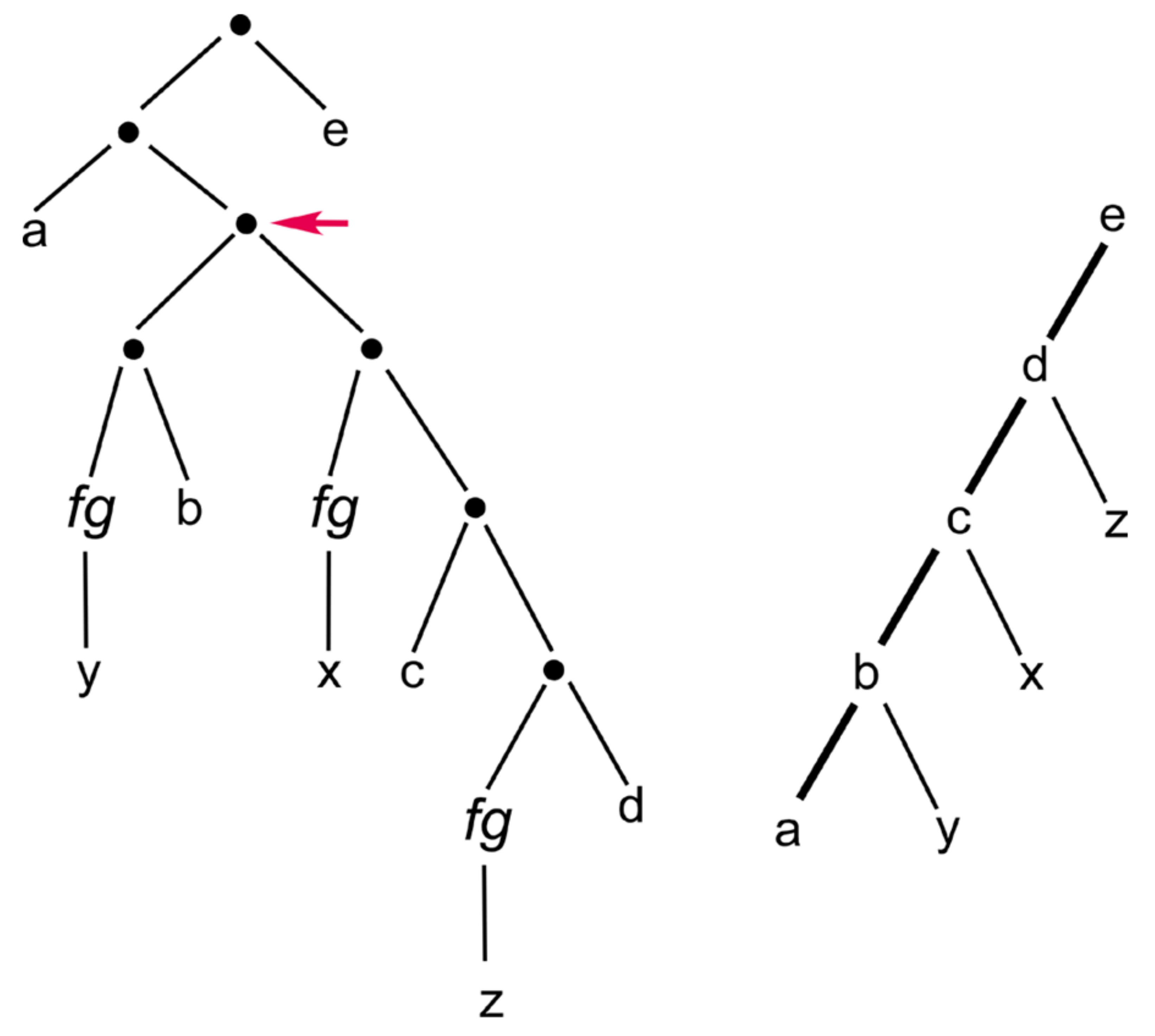}\vspace*{-1mm}
\caption{Example 4.12.}%
\end{center}\vspace*{-3mm}
\end{figure}
%EndExpansion

\bigskip
 \noindent \textbf{Example 4.12 } : Let $t=[a\bullet((\mathit{fg}(y)\bullet
b)\ \underline{\bullet}\ [\mathit{fg}(x)\bullet(c\bullet(\mathit{fg}(z)\bullet
d))])]\bullet e$ and $J:=val(t)$. See Figure 6.\ Its axis is $a<b<c<d<e$ and
$x$ defines its cut $\kappa=(\{a,b\},\{c,d,e\})$. The representative position
of $\kappa$ is the underlined occurrence of $\bullet$ marked by an arrow in
Figure 6.\ The axis of $val(t/Rep(\kappa))$ consists of $\{b,c,d\}$ with the
cut $(\{b\},\{c,d\})$. We have $\mathit{Def}(\kappa)=(\{x\},=,\{\{x\}\})$
(without axis). Nodes $y$ and $z$ are below $Rep(\kappa)$ in $t$ but are not
in $\mathit{Def}(\kappa).$\ $\square$

\medskip
Let $U\in\mathcal{U}$ and $R$ be $Rep(Cuts(U))$, defined as the set of
representing positions of the cuts of $U$. The linear order $U_{Cuts(U)}$ on
$U\uplus Cuts(U)$ is defined in Definition 2.2.

We let $U_{R}:=(U\uplus R,\leq_{in}).$ The following lemma shows that it is
isomorphic to $U_{Cuts(U)}$. The inorder $\leq_{in}$ is defined in Definition 2.17.

\bigskip
 \noindent
\textbf{Lemma 4.13} : Let $U\in\mathcal{U}$.\ If $u,v\in U$ and $\kappa
,\kappa^{\prime}\in Cuts(U),$ then : \medskip

\noindent (i) $u<v\Longleftrightarrow u<_{in}v$, \smallskip \\
(ii) $u<\kappa\Longleftrightarrow u<_{in}Rep(\kappa)$, \smallskip \\
(iii) $\kappa<u\Longleftrightarrow Rep(\kappa)<_{in}u,$  \smallskip \\
(iv) $\kappa<\kappa^{\prime}\Longleftrightarrow Rep(\kappa)<_{in}%
Rep(\kappa^{\prime}).$

 \begin{proof}  (i) has been observed in Definition 4.4(b). \medskip

(ii) and (iii). Let $u<\kappa=(U_{1},U_{2}).$ Let $w\in U_{1}$ and $z\in
U_{2}$\ be such that $Rep(\kappa)=w\sqcup_{t}z$. We have $u\in U_{1}$ hence
$u<_{in}z$ and so $u<_{t1}u\sqcup_{t}z$ and \ $z<_{t2}u\sqcup_{t}z.$\ We have
$Rep(\kappa)\leq_{t}u\sqcup_{t}z$.\ If $Rep(\kappa)=u\sqcup_{t}z$, then
$u<_{in}Rep(\kappa)$ since $u<_{t1}u\sqcup_{t}z.$ If $Rep(\kappa)<_{t}%
u\sqcup_{t}z$ then we have $Rep(\kappa)<_{t2}u\sqcup_{t}z$ as we have
$z<_{t2}u\sqcup_{t}z$ and $z<_{t}Rep(\kappa)$.\ As $u<_{t1}u\sqcup_{t}z,$ we
have $u<_{in}Rep(\kappa)$. Similarly, $\kappa<u$ implies $Rep(\kappa)<_{in}u.$

\medskip Assume now $u<_{in}Rep(\kappa)$.\ If $\kappa<u,$ then $Rep(\kappa)<_{in}u.$
Hence, we must have $x<\kappa.$

Similarly, $Rep(\kappa)<_{in}u$ implies $\kappa<u.$ \medskip

(iv) If $\kappa=(U_{1},U_{2})<\kappa^{\prime}=(U_{3},U_{4}),$ then $U_{2}$
contains some $u$ not in $U_{4}$, hence $u<U_{4}$, and thus $(U_{1}%
,U_{2})<u<(U_{3},U_{4}).$ Hence, if $\kappa<\kappa^{\prime}$, we have
$Rep(\kappa)<_{in}x<_{in}Rep(\kappa^{\prime})$.

\noindent If conversely $Rep(\kappa)<_{in}Rep(\kappa^{\prime}),$ we cannot have
$\kappa>\kappa^{\prime}$ by the previous proof, hence $\kappa<\kappa^{\prime}$.
 \end{proof}

 \noindent \textbf{Lemma 4.14}\ : (i) \ There is an MSO\ formula $\chi_{1}(x,U,W)$ such
that, if $t\in T^{\infty}(F_{\underline{M}})$, $val(t)=(N,\leq,\mathcal{U}%
,A)$, $U,W\subseteq N,x\in N$, then $\chi_{1}(x,U,W)$ holds in $\left\lfloor
t\right\rfloor $ if and only if $U\in\mathcal{U},x=Rep(U)$ and
$W=Rep(Cuts(U)).$ \medskip

(ii) Similarly, there is an MSO\ formula $\chi_{2}(x,y,U,Z)$ that holds in
$\left\lfloor t\right\rfloor $ if and only if $U\in\mathcal{U},\ x=Rep(U),$
$y=Rep(\kappa)$ where $\kappa$ is a cut of $U$, and $Z$ is the set
of\ positions $Rep(V)$ such that $V$ is an axis of an SOA-tree composing
$\mathit{Def}(\kappa)$.\medskip

(iii) There is an MSO\ formula $\chi_{3}(x,U,Z)$ that holds in $\left\lfloor
t\right\rfloor $ if and only if $U\in\mathcal{U},x=Rep(U)$ and $Z$ is the
nonempty set of representing positions $Rep(V)$ such that $V$ is an axis of an
SOA-tree composing $Tail(U)$.

 \begin{proof} (i) The following facts are MSO-expressible by the
corresponding definitions:
\begin{quote}
$U\in\mathcal{U}$, \smallskip \\
$x=Rep(U)$, \smallskip \\
$y=Rep(U_{1},U_{2})$ where $(U_{1},U_{2})$ is a cut of $U$, \smallskip \\
and $W=Rep(Cuts(U)).$
\end{quote}

From these observations, we can construct $\chi_{1}(x,U,W)$.

\medskip (ii) Similar to (i).\ An MSO formula $\alpha(U,y,X)$ can identify the set of
nodes $X$ of $\mathit{Def}(\kappa)$ where $y=Rep(\kappa)$ and $\kappa$ is a
cut of $U$.

The SO-trees composing the SO-forest $\mathit{Def}(\kappa)$\ can be
identified, and so can be their axes and thus the representing positions of
these axes.

\medskip (iii) Similar to the previous cases.
\end{proof}

\emph{Construction of a regular description scheme from a regular term.}

\medskip
We consider a regular term $t\in T^{\infty}(F_{\underline{M}})$.\ It defines a
nonempty SOA-tree $J=(N,\leq,\mathcal{U},A).$ Without loss of generality, we
assume that $A$ is not empty. Its representative position\ $Rep(A)$ \ is the root.

Our aim is to construct for $J$ a regular description scheme.\ This
construction will be based on a finite automaton accepting $t$, cf. Section 1.

Let $\mathcal{B}$ such an automaton with set of states $S$. We get a regular
term $t_{\mathcal{B}}$ (cf.\ Example 2.11(2)) by attaching to each position of
$t$, the state reached there by the unique run $run_{\mathcal{B}}$ of
$\mathcal{B}$. Each position $x$ in this term is labelled by a pair $(s,q)$
where $s$ is the symbol of $F_{\underline{M}}$ occuring at $x$ and $q\in S$.

We will construct a description scheme for $J$ of the form $\Delta
=(D,Q,d_{Ax},$\ \ $(m_{q})_{q\in Q},(w_{d})_{d\in D})$ \ where $D$ and $Q$ are
finite sets built from $S$, $m_{q}\in\mathcal{S}(D)$ for each $q\in Q$,
$w_{d}$ is a regular arrangement over $Q\cup\{\ast\}$ and $d_{Ax}\in D.$

Here is a key idea.\ Every line $U$ of the structuring of $J$\ has a
representative position $Rep(U)$ in $t$.\ The state of $\mathcal{B}$\ at
$Rep(U)$ is some $d$. If another line $U^{\prime}$ has representative position
$Rep(U^{\prime})$ with the same state $d$, then $t/Rep(U)\simeq
t/Rep(U^{\prime})$, hence $U\simeq U^{\prime}$ and even $s\triangleright
U^{+}\simeq s\triangleright U^{\prime+}$ by Lemma 4.17 below. Furthermore
$s\triangleright U^{+}$ is a regular arrangement.\ It depends only on $d$ and
is the desired $w_{d}$.\ The definition of $m_{q}$ is similar by means of
Lemma 4.16.

\medskip \noindent
\textbf{Construction 4.15\ }: \emph{From a regular term} $t$ \emph{to a
description scheme for} $val(t)$.

We let $J:=(N,\leq,\mathcal{U},A)=val(t)$ where $t$ is regular and $A$ not
empty.\ We will build a regular description scheme $\Delta=(Q,D,d_{Ax}%
,(m_{q})_{q\in Q},(w_{d})_{d\in D})$ that defines $J$.

\medskip
(i) We define the finite sets $D$ and $Q$\ :

$D:=run_{\mathcal{B}}(P_{1})\subseteq S$ where $P_{1}$ is the set of positions
$Rep(U)$ for all $U\in\mathcal{U}$.\smallskip

$Q_{tail}:=run_{\mathcal{B}}(P_{1}^{\prime})\times\{1\}$ where $P_{1}^{\prime
}$ $\subseteq P_{1}$ is the set of positions $Rep(U)$ such that $Tail(U)$ is
not empty.\smallskip

$Q_{cut}:=run_{\mathcal{B}}(P_{2})\times\{2\}$ where $P_{2}$ is the set of
positions $Rep(\kappa)$ of all cuts $\kappa\in\mathcal{K}$. \smallskip

The sets $D$ and $Q:=Q_{tail}\uplus Q_{cut}$ are finite and disjoint because
$D\subseteq S$, $Q_{tail}\subseteq S\times\{1\}$\ and $Q_{cut}\subseteq
S\times\{2\}.$ \smallskip

The element $d_{Ax}$ is $run_{\mathcal{B}}(root_{t})$ as we have
$Rep(A)=root_{t}$ since $A$ is not empty.\medskip

The multisets $m_{q}$ and the arrangements $w_{d}$ will be defined in Lemmas
4.16 and 4.17.

\medskip
(ii) We define a good labelling (Definition 3.12) :
\begin{quote}
$r(U):=run_{\mathcal{B}}(Rep(U))$ for $U\in\mathcal{U}$,
\end{quote}
\begin{quote}
$s(Tail(U)):=(run_{\mathcal{B}}(Rep(U)),1)\in Q_{tail}$ for each
$U\in\mathcal{U}$ such that $Tail(U)$ is not empty; this is well defined
because each tail $Tail(U)$ corresponds to a unique line $U$, \smallskip \\
$s(\mathit{Def}(\kappa)):=(run_{\mathcal{B}}(Rep(\kappa)),2)\in Q_{cut}$ for
each cut $\kappa\in\mathcal{K}$; this is well defined because each O-forest
$\mathit{Def}(\kappa)$ corresponds to a unique cut $\kappa$.
\end{quote}

 Furthermore, $s(\mathit{Def}(\kappa))=s(\mathit{Def}(\kappa^{\prime}))$
implies $run_{\mathcal{B}}(Rep(\kappa))=run_{\mathcal{B}}(Rep(\kappa^{\prime
}))$ hence $t/ $ \linebreak $Rep(\kappa)\simeq t/Rep(\kappa^{\prime})$ and $\mathit{Def}%
(\kappa)\simeq\mathit{Def}(\kappa^{\prime})$ by Lemma 4.11(4). Similarly,
$r(U)=r(U^{\prime})$ implies $run_{\mathcal{B}}(Rep(U))=run_{\mathcal{B}%
}(Rep(U^{\prime}))$ hence $t/Rep(U)\simeq t/Rep(U^{\prime})$ and $U\simeq
U^{\prime}$ by Lemma 4.9(2).\ Also, $s(Tail(U))=s(Tail(U^{\prime}))$ implies
$t/Rep(U)\simeq t/Rep(U^{\prime})$ and $Tail(U)\simeq Tail(U^{\prime}).$

\medskip
Lemmas 4.16 and 4.17\ will prove that we have actually defined a good
labelling and a regular description scheme.

With the previous notation, we have the following.

\bigskip
 \noindent
\textbf{Lemma 4.16 }\ :  (1) The sets $D,$ $Q_{tail}$ and $Q_{cut}$ are computable.\

\medskip (2) For each pair $(q,1)$ in $Q_{tail}$, one can compute the multiset
\begin{quote}
$m_{(q,1)}:=r\{Axes(Tail(U))\}\in\mathcal{S}(D)$,
\end{quote}
where $U\ $is\ any line such that $s(Tail(U))=(q,1).$\medskip

(3) For each pair $(q,2)$ in $Q_{cut}$, one can compute the multiset
\begin{quote}
$m_{(q,2)}:=$ \ $r\{Axes(\mathit{Def}(\kappa)\}\in\mathcal{S}(D)$,
\end{quote}
where $\kappa\ $is\ any cut such that $s(\mathit{Def}(\kappa))=(q,2).$

 \begin{proof} (1) Follows from Lemma 2.13 ($t_{\mathcal{B}}$\ is regular)
by using the MSO formulas of Lemma 4.14. \medskip

(2) \ Let $(q,1)\in Q$. One can determine the position $Rep(U)$ of some
$U\in\mathcal{U}$\ such that \\ $run_{\mathcal{B}}(Rep(U))=q.$ \smallskip

The term $t/Rep(U)$ is regular. An MSO\ formula\footnote{that does not depend
on $t$.\ } can identify the nodes $x$ of $val(t)$ (they are defined at leaves)
such that $x<U$.\ They form a set disjoint from $U$ that induces the SO-forest
$Tail(U)$. The lines $L$ in $Axes(Tail(U))$ can be identified by an MSO
formula.\ Their representing positions $Rep(L)$ are all in $t/Rep(U)$, and can
be identified by an MSO formula. For each of them, $r(L):=run_{\mathcal{B}%
}(Rep(L))$ is a state $p$ of $\mathcal{B}$\ that can be read in the symbol
$(\bullet,p)$ occuring at $Rep(L).$ (We know that an MSO\ formula determines
the symbol of $t_{\mathcal{B}}$ at each position.)\ By Lemma 2.13(2), for each
$d\in D$, the number of lines $L\in Axes(Tail(U))$ such that $r(L)=d$ can be
computed.\ Hence, we can compute $m_{(q,1)}.$\medskip

(3) The proof is similar.
\end{proof}

The next task is to prove that for each $U$\ in $\mathcal{U}$, the arrangement
$s\rhd U^{+}$ depends only on $r(U):=run_{\mathcal{B}}(Rep(U))$\ and is
regular.\ We recall that $U^{+}$ is the arrangement $(U,\leq)$ in which each
cut $\kappa(U,x)$ is inserted at its natural place and, furthermore, that is
prefixed by $\tau_{U}$\ standing for the tail of $U$ if this tail is not empty.

We will describe $s\rhd U^{+}$ as an MSO\ definable arrangement inside
$\left\lfloor t\right\rfloor $, whose linear order is $\leq_{in}$.\ For this
purpose, without introducing additional notation, we replace in $U^{+}$\ a cut
$\kappa$ by the position $Rep(\kappa)$ and we will use\ Lemma 4.13.

\bigskip
 \noindent
\textbf{Lemma 4.17} : Let $t$ be a regular term accepted by a finite automaton
$\mathcal{B}$ and $val(t)=(N,\leq,\mathcal{U},A).$ Let $U\in\mathcal{U}$.\medskip

(1) The arrangement $U^{+}$\ can be defined from $t/Rep(U)$. \smallskip

(2) The arrangement $s\rhd U^{+}$\ can be defined from $t_{\mathcal{B}%
}/Rep(U)$.\ It is regular and its MSO description can be computed.

\begin{proof}
 Let $U\in\mathcal{U}$. \medskip

(1) This is clear from definitions and previous remarks.\medskip

(2) Furthemore, $s\rhd U^{+}$\ can be defined from $t_{\mathcal{B}}%
/Rep(U)$.\ Hence, for any other line $U^{\prime}\in\mathcal{U}$, if
$r(U)=r(U^{\prime})$, we have $s\rhd U^{+}\simeq s\rhd U^{\prime+}$. \medskip

The term $t_{\mathcal{B}}/Rep(U)$ is regular as it is a subterm of the regular
term $t_{\mathcal{B}}$. We first assume that $Tail(U)$ is
empty.\ MSO\ formulas can identify the leaves belonging to $U$ and the nodes
$Rep(\kappa)$ for the cuts $\kappa$ of $U$. Let $X$ be the set of all these
nodes : we label by $\ast$ the leaves belonging to $U$ and a node
$Rep(\kappa)$ by $(run_{\mathcal{B}}(Rep(\kappa)),2)$. Then $s\rhd U^{+}%
\simeq(X,\leq_{in},lab).$\ This arrangement is regular by Proposition 2.18(2).
It is $w_{d}$\ where $d=run_{\mathcal{B}}(Rep(U))$.\ As proofs are effective,
we can build a defining MSO\ sentence describing it.\

\medskip
If $Tail(U)$ is not empty, the construction the same, except that $s\rhd
U^{+}$ is then the arrangement $(X,\leq_{in},lab)$ prefixed by $\tau_{U}$
labelled by $(run_{\mathcal{B}}(Rep(U)),1)$. Hence it is regular and is
MSO-definable.
\end{proof}

Hence Construction 4.15 and Lemmas 4.16\ and 4.17\ prove the following.

\medskip \noindent
\textbf{Proposition 4.18}: From a finite automaton $\mathcal{B}$ accepting a
term $t\in T^{\infty}(F)$ that defines an SOA-tree, one can construct a
regular description scheme for $val(t).$

\bigskip
 \noindent
\textbf{Theorem 4.19 } : The following properties of an O-tree $J$\ are equivalent:\medskip

(1) $J$ is regular,

(2) $J$ is described by a regular description scheme,

(3) $J$ is MSO$_{\mathit{fin}}$-definable.

 \begin{proof}
  \indent (1) $\Longrightarrow$(2). If $J$ is regular, a regular term
defines a structuring of it (Definition 4.4(d)).\ By Proposition 4.18 this
structuring has a regular description scheme, that describes $J$ according to
Definition 3.13.

\medskip
 (2)$\Longrightarrow$(3) If $J$ is described by a regular description scheme,
then, it is MSO$_{\mathit{fin}}$-definable by Theorem 3.18.

\medskip
 (3)$\Longrightarrow$(1).\ By Definition 4.4, the mapping $\alpha$ that
transforms the relational structure $\left\lfloor t\right\rfloor $ for $t$ in
$T^{\infty}(F)$ into the O-forest $val(t)=(N,\leq)$ (where $N=\mathrm{Occ}%
(t,\ast)$) is an MSO-transduction, because an MSO\ formula can identify the
nodes of $val(t)$ among the positions of $t$ and other formulas can define
$\leq$. \medskip

Let $J=(N,\leq)$ be an MSO$_{\mathit{fin}}$-definable O-tree.\ It is, up to
isomorphism, the unique model of an MSO$_{\mathit{fin}}$ sentence $\psi$.\ It
follows by a standard argument\footnote{If $\alpha$ is an MSO-transduction and
$\psi$ \ is an MSO$_{fin}$ sentence, then the set of structures $S$ such that
$\alpha(S)\models\psi$ \ is MSO$_{fin}$-definable, by Backwards Translation
\cite{CouEng}.} that the set\ of terms $t$ in $T^{\infty}(F)$ such that
$\alpha(\left\lfloor t\right\rfloor )\models\psi$ is MSO$_{\mathit{fin}}$
definable.\ Since the structures $\left\lfloor t\right\rfloor $ are linearly
ordered by $\leq_{in}$ that is MSO-definable, this set is also MSO-definable
(cf.\ Definition 2.8), and thus, contains a regular term, by a result due to
Rabin \cite{Tho90}.\ This term defines $J$.\ Hence $J$ is regular.
\end{proof}

As for Corollaries 4.22 and 4.31\ in \cite{CouLMCS}\ we have :

\bigskip
 \noindent\textbf{Corollary 4.20 } : The isomorphism problem for regular O-trees is decidable.

 \begin{proof}
  A regular O-tree can be defined by a regular term or by an
MSO$_{fin}$ sentence. The proof of Theorem 4.19\ is effective: algorithms can
convert any of these specifications into another one.\ Hence, two regular
O-trees can be given, one by an MSO$_{fin}$\ sentence $\psi$, the other by a
regular term $t$.\ They are isomorphic if and only if $\alpha(\left\lfloor
t\right\rfloor )\models\psi$ (cf.\ the proof of (3)$\Longrightarrow$(1) of
Theorem 4.19) if and only if $\left\lfloor t\right\rfloor \models\psi^{\prime
}$ where $\psi^{\prime}$ obtained by applying Backwards Translation
\cite{CouEng} to the sentence $\psi$ and the transduction $\alpha$. This is
decidable \cite{Tho90}.
\end{proof}

\smallskip\noindent
\textbf{Corollary 4.21} : An O-forest is regular if and only if it is
MSO$_{\mathit{fin}}$-definable. The isomorphism of two regular O-forests is decidable.

 \begin{proof}
  If $J$ is an O-forest, then $J\bullet\ast$ is an O-tree. It is
easy to prove that $J$\ is MSO$_{\mathit{fin}}$-definable if and only if
$J\bullet\ast$ is.\ Furthermore, $J$ is regular if and only if $J\bullet\ast$
is.\ The results follow then from Theorem 4.19\ and Corollary
4.20.
 \end{proof}

\bigskip
 \noindent
\textbf{Corollary 4.22}\ : (1) The MSO$_{\mathit{fin}}$-theory of a regular
O-forest is decidable.\medskip

(2) It is decidable whether some O-forest satisfies a given MSO$_{\mathit{fin}%
}$-sentence.\ If so, this sentence is satisfied by some regular O-forest.

 \begin{proof}
  We use routine constructions as for (3)$\Longrightarrow$(1)
of Theorem 4.19.
 \end{proof}

\section{Conclusion}

We have shown how to describe O-trees and O-forests, \emph{u.t.i}., by
infinite terms over three operations. We have generalized to regular O-forests
the results of \cite{CouLMCS} that concern regular join-trees.\

More complex terms than the regular ones have finitary descriptions (see
\cite{Blu}) and thus, can yield finitary descriptions of other types of
O-forests. As they have decidable MSO theories, the MSO-theories of the
defined O-forests are decidable.

\subsection*{Acknowledgements} I thank the organizers of the conference dedicated
to the memory of B.\ Trakhtenbrot and the referee for his or her helpful comments.

\end{document}